\documentclass[twocolumn,amsmath,amssymb,prd,reprint,longbibliography,floatfix,8pt]{revtex4-1}

\usepackage{cancel}
\usepackage{enumitem}
\usepackage[utf8x]{inputenc}
\usepackage{graphicx}
\usepackage{dcolumn}
\usepackage{bm}
\usepackage{hyperref}
\usepackage[switch]{lineno}
\usepackage{float}             
\usepackage{subfigure}
\usepackage{tikz}
\usepackage{hyperref}
\usepackage{tabularx}
\usepackage{adjustbox}
\usepackage{multirow}
\usepackage{xr}
\usetikzlibrary{arrows.meta}

\begin{document}

\title{Unravelling the contributions to spin-lattice relaxation in Kramers single-molecule magnets}

\author{Sourav Mondal}
\author{Alessandro Lunghi}
\email{lunghia@tcd.ie}
\affiliation{School of Physics, AMBER and CRANN Institute, Trinity College, Dublin 2, Ireland}

\begin{abstract}
{\bf The study of how spin interacts with lattice vibrations and relaxes to equilibrium provides unique insights on its chemical environment and the relation between electronic structure and molecular composition. Despite its importance for several disciplines, ranging from magnetic resonance to quantum technologies, a convincing interpretation of spin dynamics in crystals of magnetic molecules is still lacking due to the challenging experimental determination of the correct spin relaxation mechanism. We apply \textit{ab initio} spin dynamics to a series of twelve coordination complexes of Co$^{2+}$ and Dy$^{3+}$ ions selected among $\sim$240 compounds that largely cover the literature on single-molecule magnets and well represent different regimes of spin relaxation. Simulations reveal that the Orbach spin relaxation rate of known compounds mostly depends on the ions' zero-field splitting and little on the details of molecular vibrations. Raman relaxation is instead found to be also significantly affected by the features of low-energy phonons. These results provide a complete understanding of the factors limiting spin lifetime in single-molecule magnets and revisit years of experimental investigations by making it possible to transparently distinguish Orbach and Raman relaxation mechanisms.}
\end{abstract}

\maketitle

\section*{Introduction}

Coordination compounds of first-row transition metals and lanthanide ions offer a vast playground for the exploration of electronic and magnetic properties for applications ranging from catalysis\cite{gupta2008catalytic} and sensors\cite{cable2011luminescent} to luminescence\cite{buenzli2015design,wegeberg2021luminescent}. In particular, molecules showing magnetic properties due to the presence of unpaired $d$/$f$ electrons are under an intense scrutiny for applications in the areas of information storage\cite{sessoli1993magnetic}, spintronics\cite{rocha2006spin} and quantum science\cite{atzori2019second,gaita2019molecular,wasielewski2020exploiting}. However, the delivery of molecule-based technologies strongly relies on the possibility to overcome their short spin lifetime. Similarly to hard ferromagnets\cite{coey2002permanent}, coordination compounds possessing easy-axis magnetic anisotropy are known to exhibit long spin relaxation time, hence their name single-molecule magnets (SMMs)\cite{coronado2020molecular}. Unless cryogenic temperatures are achieved, the interaction between spin and phonons, namely the spin-phonon coupling, is the main responsible for magnetic moment relaxation\cite{lunghi2022toward}. Early studies \cite{sessoli1993magnetic,accorsi2006tuning} have shown that spin relaxation time, $\tau$, of SMMs follows an Arrhenius-like law 
\begin{equation}
\tau=\tau_0 e^{U_{eff}/k_B T}\:,
\label{Arr}
\end{equation}
where for a given temperature, $T$, the pre-exponential factor $\tau_0$ set the relaxation time-scale and the $U_{eff}$ represents an effective magnetic moment reversal barrier due to the presence of magnetic anisotropy. $U_{eff}$ is intimately connected to the electronic structure of magnetic ions and it coincides with the energy of the electronic excited state promoting relaxation through the absorption and emission of a series of phonons\cite{lunghi2022toward}, \textit{i.e.} the Orbach relaxation mechanism. Increasing $U_{eff}$ has been the main strategy to improve $\tau$ and many efforts have been devoted to engineering coordination compounds with large zero-field splittings\cite{accorsi2006tuning,rinehart2011exploiting}. The most successful strategy employs the use of Co$^{2+}$ or Dy$^{3+}$ ions with a strong and axial crystal field as building blocks for SMMs\cite{zabala2021single}. Record values of 450\cite{bunting2018linear} and 1541\cite{guo2018magnetic} have been reached for single-ion complexes of Co$^{2+}$ and Dy$^{3+}$, respectively. 

Now that the limit in crystal field axiality has been virtually reached\cite{guo2018magnetic}, new strategies towards high-temperature SMMs are required. Coupling multiple ions stands as one possible route and important milestones in this direction have been achieved\cite{rinehart2011strong,gould2022ultrahard,magott2022intermetallic}. Another strategy instead requires to look at the entire spin-phonon relaxation process to determine other physical quantities that influence it. Here we pursue the latter approach.

\begin{figure*}[t]
    \centering
    \includegraphics[scale=1]{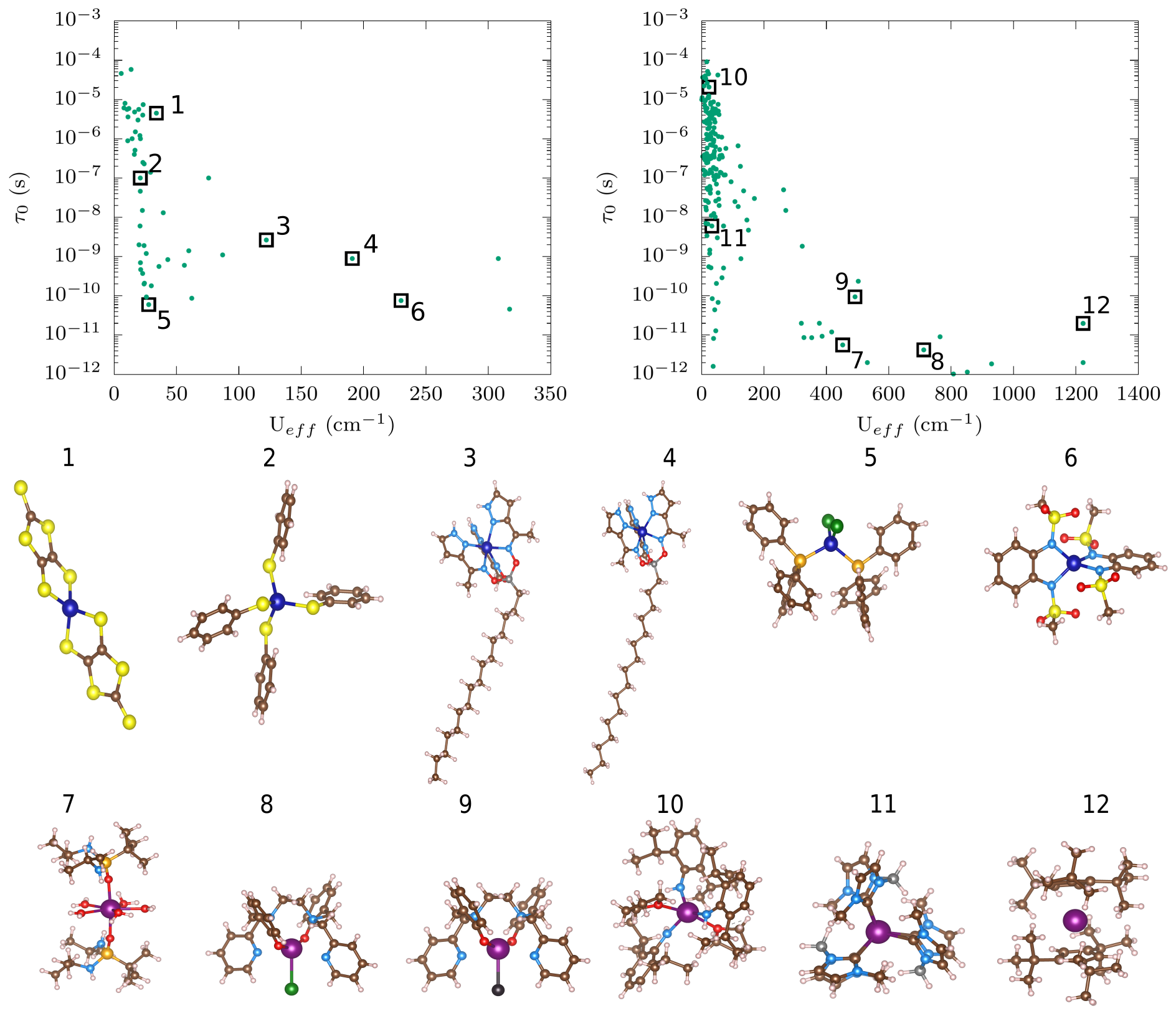}
    \caption{\textbf{Experimental correlations and molecular structures.} The top left and top right panels report the experimental $\tau_0$ vs $U_{eff}$ for a set of $\sim$240 single-ion Co$^{2+}$ and Dy$^{3+}$ complexes, respectively, individuated by scraping the literature on single-molecule magnets or from the SIMDAVIS database\cite{duan2021data}. Black squares are used to identify the twelve molecules selected for the study. The middle panel reports the molecular geometries of selected six Co$^{2+}$ complexes \textbf{1-6}. The bottom panel reports the molecular geometries of selected six Dy$^{3+}$ complexes \textbf{7-12}. Color codes for atoms: Dy in purple, Co in indigo, N in light blue, O in red, B in gray, Br in dark green, C in dark brown, S in yellow, P in gold, Cl in black, H in pale pink. }
    \label{image1}
    \hfill
\end{figure*}

A striking example of how our limited knowledge of spin-phonon relaxation has impacted the field of SMMs comes from the visualization of the correlation between $\tau_0$ and $U_{eff}$. 
The top panel of Fig. \ref{image1} reports these two quantities for $\sim$240 single-ion Co$^{2+}$ and Dy$^{3+}$ SMMs that largely cover the relevant literature until early 2019 and highlights how $\tau_0$ spans eight orders of magnitude and strongly correlates with $U_{eff}$, undercutting the large values achieved for the latter. This is a result of the fact that despite the many efforts to control $U_{eff}$, no clear insight on how to chemically control the pre-exponential factor $\tau_0$ is yet available and no attempts in optimizing it has ever been made. Moreover, as values of $U_{eff}$ above 30 K had been reported, strong deviations from Eq. \ref{Arr} have been observed and attributed to Raman relaxation\cite{harman2010slow,zadrozny2011slow}, \textit{i.e.} a process involving spin transitions due to the simultaneous absorption and emission of two phonons. Only recently theoretical models have been able to shine some light on the nature of this mechanism, showing that Raman mechanisms can also present an Arrhenius like behaviour under certain conditions\cite{lunghi2020multiple,gu2020origins,briganti2021complete}. This situation poses an important challenge. Although experimental and computational strategies have been designed to disentangle Orbach and Raman relaxation\cite{pedersen2015design,rechkemmer2016four}, discerning the two mechanisms is far from trivial and misinterpretation of the fitted parameters in Eq. \ref{Arr} has been suggested\cite{pedersen2015design,novikov2015trigonal,rechkemmer2015comprehensive,rechkemmer2016four,lunghi2017role,gu2020origins}. Understanding the nature of these contributions to spin relaxation and their dependency from chemical structure is a fundamental step toward controlling spin-phonon coupling and deliver improved SMMs. 

Here we aim at providing a deeper understanding of the contributions to spin-phonon relaxation in single-molecule magnets with the goal of removing ambiguities in the interpretation of experiments and conclusively establish what determines the rate of relaxation time. To achieve this we exploit our recently developed \textit{ab initio} spin dynamics approach, where the time evolution of the molecular magnetic moment under the influence of phonons is predicted from first principles and without the need of any information from the experiments except for the crystal structure\cite{lunghi2017intra,lunghi2019phonons,lunghi2020multiple,briganti2021complete,lunghi2022toward}. It has been demonstrated that this method can quantitatively describe both one- and two-phonon processes, responsible for Orbach and Raman relaxation mechanisms, across the entire relevant temperature range\cite{lunghi2022toward}. Here we apply this strategy to a total of twelve Co$^{2+}$ and Dy$^{3+}$ SMMs. These compounds are selected from all the $\sim$ 240 single-ion SMMs identified in literature\cite{duan2021data} to sample different regimes of $\tau_0$ vs $U_{eff}$ and provide an unprecedented benchmark for both simulations and experiments.

We show that the correlation between $\tau_0$ and $U_{eff}$ arising from literature is only virtual and due to years of misinterpretation of Raman relaxation mechanism as Orbach. Thanks to the access to all the details of spin relaxation, we demonstrate that the variance in the Orbach relaxation time among different SMMs is largely determined by the static crystal field splitting, while Raman relaxation time is also dependent on the details of molecular vibrations and spin-phonon coupling, providing a revised road map for the design of improved SMMs.

\section*{Computational Methods}

{\bf Molecules selection.} The SIMDAVIS database has been used to gather a set of 183 Dy$^{3+}$ single-ion SMMs and their respective relaxation data\cite{duan2021data}. A manual search for the corresponding Co$^{2+}$ single-ion SMMs published until early 2019 has instead been carried out and resulted in 56 compounds. Table S1 reports all the references, $U_{eff}$ and $\tau_0$ for the Cobalt SMMs. Although our analysis does not account for the literature in its entirety, the selected compounds represent the entire range of relaxation regimes. Twelve molecules were selected from this data set following these criteria as closely as possible: i) reported relaxation data span the entire range of $\tau_0$ vs $U_{eff}$, ii) molecules are chemically and structurally diverse. The six Co$^{2+}$ and six Dy$^{3+}$ molecules chosen for this study are [Co(C$_3$S$_5$)$_2$](Ph$_4$P)$_2$\cite{fataftah2014mononuclear} (\textbf{1}), [Co(SPh)$_4$](Ph$_4$P)\cite{zadrozny2011slow} (\textbf{2}), $\beta$-Co\cite{pavlov2016polymorphism} (\textbf{3}), $\alpha$-Co\cite{pavlov2016polymorphism} (\textbf{4}), [Co(PPh$_3$)$_2$Br$_2$]\cite{SMM5} (\textbf{5}), [CoL$_2$][ (HNEt$_3$)$_2$]\cite{SMM6}, where H$_2$L= 1,2-bis(methane-sulfonamido)benzene (\textbf{6}), [L$_2$Dy(H$_2$O)$_5$][I]$_3\cdot$L$_2\cdot$H$_2$O\cite{SMM7} (\textbf{7}) where L= $^t$BuPO(NH$^i$Pr)$_2$, [Dy(bbpen)Br]\cite{SMM8} (\textbf{8}) where H$_2$bbpen= N,N'-bis(2-hydroxybenzyl)-N,N'-bis(2-methylpyridyl)ethylenediamine), [Dy(bbpen)Cl]\cite{SMM8} (\textbf{9}), Dy[NHPh$^i$Pr$_2$]$_3$(THF)$_2$\cite{SMM10} (\textbf{10}), Dy(Bc$^\text{Me}$)$_3$\cite{SMM11} (\textbf{11})  where [Bc$^{\text{Me}}$]$^{−}$ = dihydrobis(methylimidazolyl)borate, and Dy[Cp$_2^\text{ttt}$][B(C$_6$F$_5$)$_4$]\cite{SMM12} (\textbf{12}) where (Cp$^\text{ttt}$ = C$_5$H$_5$Bu$_3$-1,2,4). The chemical structure of \textbf{1}-\textbf{12} is reported in the bottom panel of Fig. \ref{image1}. Compounds \textbf{1}-\textbf{12} all respect the criteria i)-ii) except for the pairs \textbf{3}-\textbf{4} and \textbf{8}-\textbf{9}, which have instead been chosen to challenge \textit{ab initio} spin dynamics over  minimal structural variations.\\

{\bf Electronic structure simulations.} Cell and geometry optimization and simulations of $\Gamma$-point phonons have been performed with periodic density functional theory (pDFT) using the software CP2K.\cite{kuhne2020cp2k} Cell optimization was performed employing a very tight force convergence criteria of 10$^{−7}$ a.u. and SCF convergence criteria of 10$^{-10}$ a.u. for the energy. A plane wave cutoff of 1000 Ry, DZVP-MOLOPT Gaussian basis sets, and Goedecker-Tetter Hutter pseudopotentials\cite{goedecker1996separable} were employed for all atoms. The  Perdew-Burke-Ernzerhof (PBE) functional and DFT-D3 dispersion corrections were used.\cite{perdew1996generalized,grimme2010consistent}. 

ORCA\cite{neese2020orca}  had been used to compute the magnetic properties. Magnetic properties for Dy$^{3+}$ ions were computed from the CASSCF calculations employing active space of seven 4$f$ orbitals with nine electrons (9,7) and by using all the solutions with multiplicity six, 224 solutions with multiplicity four, and 490 solutions with multiplicity two. Similarly, for Co$^{2+}$ ions, magnetic properties were computed from the CASSCF calculations employing active space of five 3$d$ orbitals with seven electrons (7,5) and by using all the solutions with multiplicity four, and 40 solutions with multiplicity two. The RIJCOSX approximation for coulomb integral and the integration grid of GridX6 were used for both the ions. The basis sets DKH-def2-QZVPP for Co atoms, DKH-def2-SVP for H and SARC2-DKH-QZVP for Dy atoms were used. DKH-def2-TZVPP basis set has been used for the rest of the atoms present in the systems. \\

{\bf Spin-phonon coupling and relaxation simulations.} First order spin-phonon coupling coefficients $(\partial \hat{H}_{s}/\partial Q_{\alpha})$ are computed as
\begin{equation}
    \left( \frac{\partial \hat{H}_{s}}{\partial Q_{{\alpha}}} \right )= \sum_{i}^{3N} \sqrt{\frac{\hbar}{2\omega_{\alpha}m_{i}}} L_{\alpha i}\left( \frac{\partial \hat{H}_{s}}{\partial X_i} \right )\:.
    \label{num_diff}
\end{equation}
where $Q_{{\alpha}}$ is the displacement vector associated with the $\alpha$-phonon and $N$ is the number of atoms in the unit cell. $L_{\alpha i}$ and $\omega_{\alpha}$ are the Hessian matrix eigenvectors and the phonons angular frequency. Only $\Gamma$-point phonons are used. The first order derivatives of the spin Hamiltonian with respect to the Cartesian degree of freedom $X_i$, $(\partial \hat{H}_{s} / \partial X_i)$, is computed by numerical differentiation\cite{lunghi2017intra}. Each molecular degree of freedom is sampled eight times between $\pm$ 0.08 \AA$ $. Spin-phonon coupling coefficients are used to calculate the spin-phonon relaxation time on the basis of Redfield equations.\cite{lunghi2019phonons,lunghi2020multiple,lunghi2022toward} First- and second-order time-dependent perturbation theory have been used to simulate both one- and two-phonon processes. The software MolForge is used for these simulations and it is freely available at github.com/LunghiGroup/MolForge\cite{lunghi2022toward}. As discussed elsewhere, the simulation of Kramers systems in zero external field requires the use of the non-diagonal secular approximation, where population and coherence terms of the density matrix are not independent from one another. This is achieved by simulating the dynamics of the entire density matrix for one-phonon processes\cite{lunghi2019phonons,lunghi2022toward}. An equation that accounts for the dynamics of the entire density matrix under the effect of two-phonon processes resulting from fourth-order time dependent perturbation theory is not yet available. However, it is possible to remove the coupling between population and coherence terms by orienting the molecular easy axis along the quantization $z$-axis and by applying a small magnetic field to break Kramers degeneracy\cite{lunghi2022toward}. Here we employ the latter strategy to simulate Raman relaxation. 

\vspace{0.5cm}

\section*{Results}

\subsection*{\textit{Ab Initio} Spin Dynamics}

The bottom panel of Fig.~\ref{image1} shows the molecular structure of the six Co$^{2+}$ and six Dy$^{3+}$ molecules chosen for this study. The corresponding values of $\tau$ and $U_{eff}$ are highlighted in the top panel of Fig. \ref{image1}, showing the large span of values of these quantities. Compounds \textbf{6} and \textbf{12} are among the SMMs with the highest values of $U_{eff}$, while \textbf{1} and \textbf{10} are reported to have a very large $\tau_0$ and small $U_{eff}$. These compounds also show a varied coordination chemistry. For instance, \textbf{1}, \textbf{2}, \textbf{5}, and \textbf{6} are possess a tetrahedrally coordinated Co$^{2+}$ ion, while \textbf{3} and \textbf{4} show octahedral coordination. Similarly, molecules \textbf{10} and \textbf{11} among the Dy$^{3+}$ complexes show coordination number of six and molecule \textbf{7} shows the coordination number of seven. On the other hand, compounds \textbf{8}-\textbf{9}, and \textbf{12} show the coordination number of three and bis-$\eta^{5}$, respectively. The chosen compounds also show a varied set of ligands and charge states. The Co$^{2+}$ complexes present a metal ion coordinated by N or S donor atoms in most cases, except in \textbf{5} where Co is bonded by Br and P donor atoms. Dy$^{3+}$ also present a varied coordination, including Oxygen, Nitrogen, halide ions and metal-organic bonds. In most cases, the molecular unit is charged, except for \textbf{5} which is neutral. For instance, \textbf{1}, \textbf{2}, \textbf{6} are di-negative anions and \textbf{3} and \textbf{4} are mono-positive cations. On the contrary, most of the Dy complexes are neutral, except \textbf{7} and \textbf{12}, which are both cations.

With exception of \textbf{6} and \textbf{12} that were studied in a previous work \cite{lunghi2020multiple,lunghi2022toward}, all the unit cells of these compounds were optimized with pDFT, as described in the Computational Methods section.
Electronic structure simulations at the level of CASSCF are then carried out on all isolated molecular structures with the coordinates fixed to the pDFT optimized value. The magnetic properties of all the compounds are then computed by mapping electronic structure results onto effective Hamiltonians. The effective spin Hamiltonian 
\begin{equation}
    \hat{H}_{S}= D \hat{S}^{2}_{z} + E \left ( \hat{S}^{2}_{x} -\hat{S}^{2}_{y} \right )\:,
    \label{sHam}
\end{equation}
is used to describe the ground state of all Co$^{2+}$ compounds, while an effective Crystal Field Hamiltonian is used for Dy$^{3+}$ compounds
\begin{equation}
    \hat{H}_{CF}= \sum_{l=2,4,6} \sum_{m=-l}^{l} B_{m}^{l} \hat{O}_{m}^{l}\:,
    \label{CFHam}
\end{equation}
where the operators $\hat{O}_{m}^{l}$ are tesseral function of the total angular momentum operators, $\vec{\mathbf{J}}$. We will refer to any of the two operators with $\hat{H}_{0}$ in the following. All the studied Co$^{2+}$ compounds are found to exhibit a $S=3/2$ ground state with uni-axial anisotropy ($D<0$), while the Dy$^{3+}$ compounds show a $M_j=\pm 15/2$ ground-state Kramers doublet (KD) well separated in energy from the excited KDs, see Table \ref{T1} and Table S4 in ESI.  %
\begin{table}[h!]
    \centering
    \caption{ \textbf{Simulated energy data for the Kramer's Doublets.}
    \textbf{$\Delta_{01}$} represents the difference of energy between the first excited state and the ground state  Kramers doublets, and \textbf{$\Delta_{07}$}  represents the difference of energy between the last and the ground state Kramers doublets in Dy$^{3+}$ compounds.} 
\begin{tabular}{|c|c|c|}
   \hline
        System &  \textbf{$\Delta_{01}$} (cm$^{-1}$)  & \textbf{$\Delta_{07}$} (cm$^{-1}$) \\ 
        \hline
  \textbf{1}  & 273.51 & - \\
  \textbf{2}  & 93.44 & -  \\
  \textbf{3}  & 199.56 & - \\
  \textbf{4}  & 224.78 & - \\
  \textbf{5}  & 20.06 & -  \\
  \textbf{6} & 198.08 & -\\
  \textbf{7} & 293.84 & 706.85 \\
  \textbf{8} & 394.41 & 864.21 \\
  \textbf{9} & 370.51 & 810.70 \\
  \textbf{10} & 240.94  & 968.38\\
  \textbf{11} & 10.55  & 520.95 \\
  \textbf{12} & 451.00 & 1473.01\\
        \hline
    \end{tabular}
    \label{T1}
\end{table} 

\begin{figure*}[!htbp]
    \centering
    \includegraphics[scale=1]{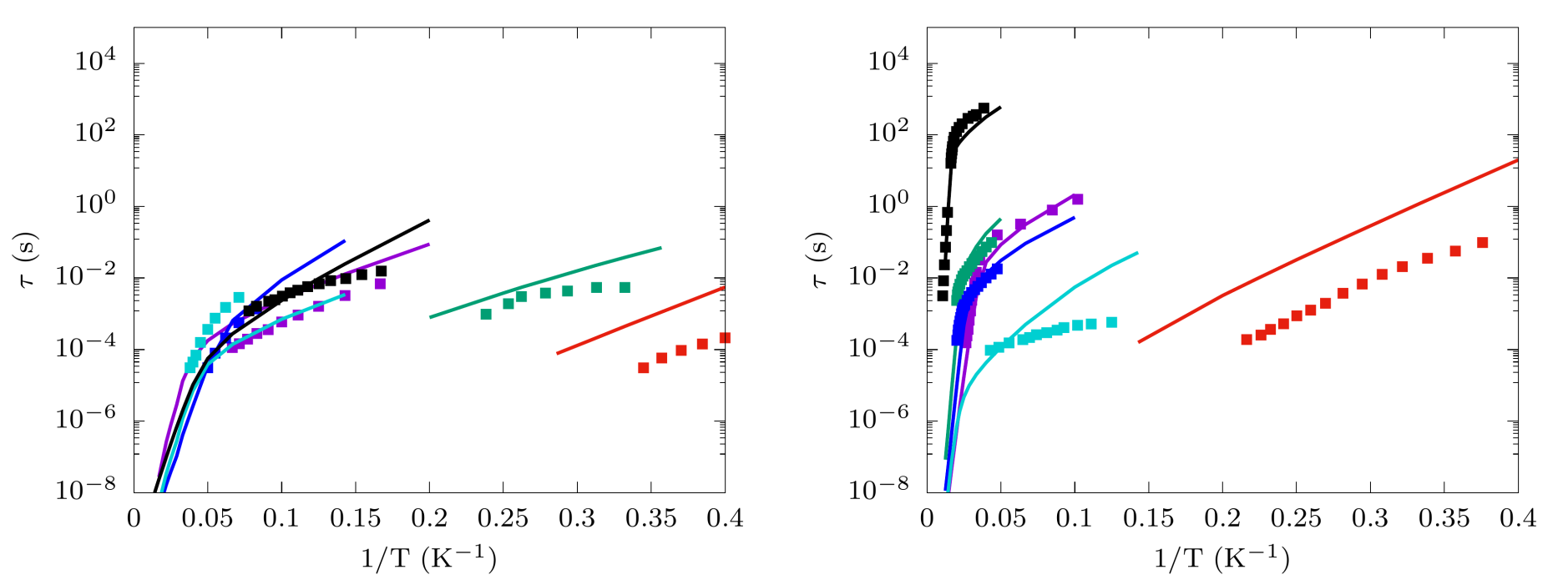}
    \caption{\textbf{Spin-phonon relaxation times.} Simulated values of $\tau$ are reported with continuous lines, while experimental values are reported with solid square symbols. Color code for left panel: 1 (violet), 2 (green), 3 (blue), 4 (turquoise), 5 (red), and 6 (black). Color code for right panel: 7 (violet), 8 (green), 9 (blue), 10 (turquoise), 11 (red), and 12 (black).} 
    \label{image2}
    \hfill
\end{figure*}


Once the eigenstates, $| a \rangle$, and eigenvalues, $E_{a}$, of these operators have been obtained, spin dynamics can be simulated by computing the transition rate among different spin states, $W_{ab}$. Spin relaxation in molecular Kramers systems with large magnetic anisotropy takes contributions from one- and two-phonon processes. Considering one-phonon processes, the transition rate, $\hat{W}^{1-ph}_{ba}$, among spin states reads 
\begin{equation}
    \hat{W}^{1-ph}_{ba}=\frac{2\pi}{\hbar^2} \sum_{\alpha} | \langle b| \left(\frac{\partial \hat{H}_{0}}{\partial Q_{\alpha}} \right) |a \rangle  |^2G^{1-ph}(\omega_{ba}, \omega_{\alpha}) \:, 
    \label{Orbach}
\end{equation}
where $\hbar\omega_{ba}=E_{b}-E_{a}$ and the term $\left( \partial \hat{H}_{0} / \partial Q_{\alpha} \right)$ provides the intensity of the coupling between spin and the $\alpha$-phonon $Q_{\alpha}$. The function $G^{1-ph}$ reads
\begin{equation}
G^{1-Ph}(\omega, \omega_{\alpha}) = \delta(\omega-\omega_\alpha)\bar{n}_\alpha +\delta(\omega +\omega_\alpha)(\bar{n}_\alpha +1)    \:,
\end{equation}
where $ \bar{n}_\alpha=(e^{\hbar\omega_{\alpha}/k_BT} -1)^{-1}$ is the Bose−Einstein distribution accounting for the phonons' thermal population, $k_B$ is the Boltzmann constant, and the Dirac delta functions enforce energy conservation during the absorption and emission of phonon by the spin system, respectively. Eq. \ref{Orbach} accounts for the Orbach relaxation mechanism, where a series of phonon absorption processes leads the spin from the fully polarized state $M_{s}=S$ to an excited state with an intermediate value of $M_{s}$ before the spin can emit phonons back to $M_{s}=-S$, and similarly for states characterized by the total angular momentum, $J$. 

Two-phonon processes provide an alternative pathway of relaxation to equilibrium, namely the Raman mechanism. We model two-phonon spin-phonon transitions, $W^{2-ph}_{ba}$, as
\begin{equation}\label{Raman}
    \hat{W}^{2-ph}_{ba}  =\frac{2\pi}{\hbar^2} \sum_{\alpha\beta}\left | T^{\alpha\beta,+}_{ab} + T^{\beta\alpha,-}_{ab} \right|^2G^{2-ph} (\omega_{ba}, \omega_{\alpha}, \omega_{\beta})\:,
\end{equation}
where the terms
\begin{equation}
T^{\alpha\beta,\pm}_{ab} = \sum_{c} \frac{ \langle a| (\partial \hat{H}_{s}/\partial Q_{{\alpha}}) |c\rangle \langle c| (\partial \hat{H}_{s}/\partial Q_{{\beta}})|b\rangle }{E_c -E_b \pm \hbar\omega_\beta} 
\end{equation}
involve the contribution of all the spin states $|c\rangle $ at the same time, often referred to as a virtual state. The function $G^{2-ph}$ fulfills a similar role as $G^{1-ph}$ for one-phonon processes, and includes contributions from the Bose-Einstein distribution and imposes energy conservation. $G^{2-ph}$ accounts for all two-phonon processes, i.e. absorption of two phonons, emission of two-phonons or absorption of one phonon and emission of a second one. The latter process is the one that determines Raman relaxation rate, and in this case $G^{2-ph}$ reads
\begin{equation}
G^{2-ph}(\omega, \omega_{\alpha},\omega_{\beta}) = \delta(\omega-\omega_\alpha+\omega_\beta)\hat{n}_\alpha(\hat{n}_\beta +1).
\end{equation}

All the parameters appearing in Eqs.~\ref{Orbach} and ~\ref{Raman} are computed from first principles (see computational methods section). In a nutshel, lattice harmonic frequencies, $\omega_{\alpha}/2\pi$, and normal modes, $Q_{\alpha}$, are computed by finite differentiation after geometry optimization with pDFT. 
All the parameters appearing in Eqs. \ref{sHam} and \ref{CFHam}, \textit{i.e.} $D$, $E$, and  $B^{l}_{m}$, are numerically differentiated with respect to the atomic displacements defined by $Q_{\alpha}$ to obtain the spin-phonon coupling coefficients $\left( \partial \hat{H}_{s} / \partial Q_{\alpha} \right)$. Once all the matrix elements $W^{n-ph}_{ba}$ have been computed, $\tau^{-1}$ can be predicted by simply diagonalizing $W^{n-ph}_{ba}$ and taking the smallest non-zero eigenvalue. The study of $W^{1-ph}$ provides the Orbach contribution to the relaxation rate, $\tau^{-1}_{Orbach}$, while $W^{2-ph}$ provides the Raman contribution, $\tau^{-1}_{Raman}$. The total relaxation time is thus computed as $\tau^{-1}= \tau^{-1}_{Orbach} + \tau^{-1}_{Raman}$.

Fig.~\ref{image2} reports the experimental values and the simulations results for $\tau$ as a function of temperature for both Co$^{2+}$ and Dy$^{3+}$ complexes. Overall simulations reproduce experimental results very well and they prove capable of reproducing trends in relaxation rate for different molecules without any input from experiments nor adjustable parameters in their equations. The best results are obtained for \textbf{3}, \textbf{6}, \textbf{7}, \textbf{8}, \textbf{9}, and \textbf{12} where the deviation between experiments and simulations is vanishingly small. In particular, the comparison between \textbf{8} and \textbf{9} is illustrative of the power of \textit{ab initio} simulations, which are shown to be able to predict differences in relaxation times coming from substituting a Cl$^{-}$ with a Br$^{-}$ as ligands in the first coordination sphere. Interestingly, in the case of Co$^{2+}$ compounds with large $U_{eff}$, such as for \textbf{6}, \textbf{4}, and \textbf{3}, we predict values of $\tau$ that fall within the same order of magnitude. Although simulations are not able to perfectly distinguish different Co$^{2+}$ molecules to this degree of accuracy, the relaxation time is in good agreement with experimental observations. It is important to note that the contribution to relaxation coming from dipolar-mediated cross-relaxation, not included in simulations, is an important factor that potentially contributes to the residual deviations between simulations and experiments. Experimental results for diluted compounds or in external field are often not available (see Table S6) and for instance, the effect of dipolar relaxation is particular visible in \textbf{2}, where experimental relaxation times start flattening out at low temperature. The importance of accounting for dipolar cross-relaxation in the comparison between relaxation data and simulations has been recognized for $S=1/2$ systems\cite{lunghi2022toward} and supports the hypothesis that the residual errors for Co$^{2+}$ SMMs is due to this effect. This argument is also in agreement with the higher accuracy obtained in high-$U_{eff}$ Dy SMMs, which are naturally screened from dipolar relaxation. The largest deviations are observed for \textbf{5} and \textbf{11}. These compounds have the smallest zero-field splittings and we attribute these somewhat larger errors to the absence of acoustic and border-zone phonons in our simulations. The latter are not accounted for by simulating the sole unit-cell phonons and their absence mostly affects the low-energy vibrational density of states in resonance with the spin transitions of \textbf{5} and \textbf{11}. Finally, we note that the simulation of relaxation times in Dy SMMs is particularly sensitive to the accuracy of the crystal field parameters appearing in $\hat{H}_{CF}$ and thus to the quality of the modellization of the magnetic ion's coordination sphere. Geometries optimized with pDFT and the use of the first coordination sphere for the simulation of $\hat{H}_{CF}$ were found to be accurate in all cases except for \textbf{7}, where the Dy-H$_2$O distances are not well reproduced by DFT and the inclusion of the second coordination sphere is necessary. Accurate results were obtained by simulating $\hat{H}_{CF}$ for a model including first and second coordination spheres with experimental X-ray distances (see Fig. S6). 

\subsection*{Analysis of Spin Relaxation}

\begin{figure}[!h]
    \centering
    \includegraphics[scale=1]{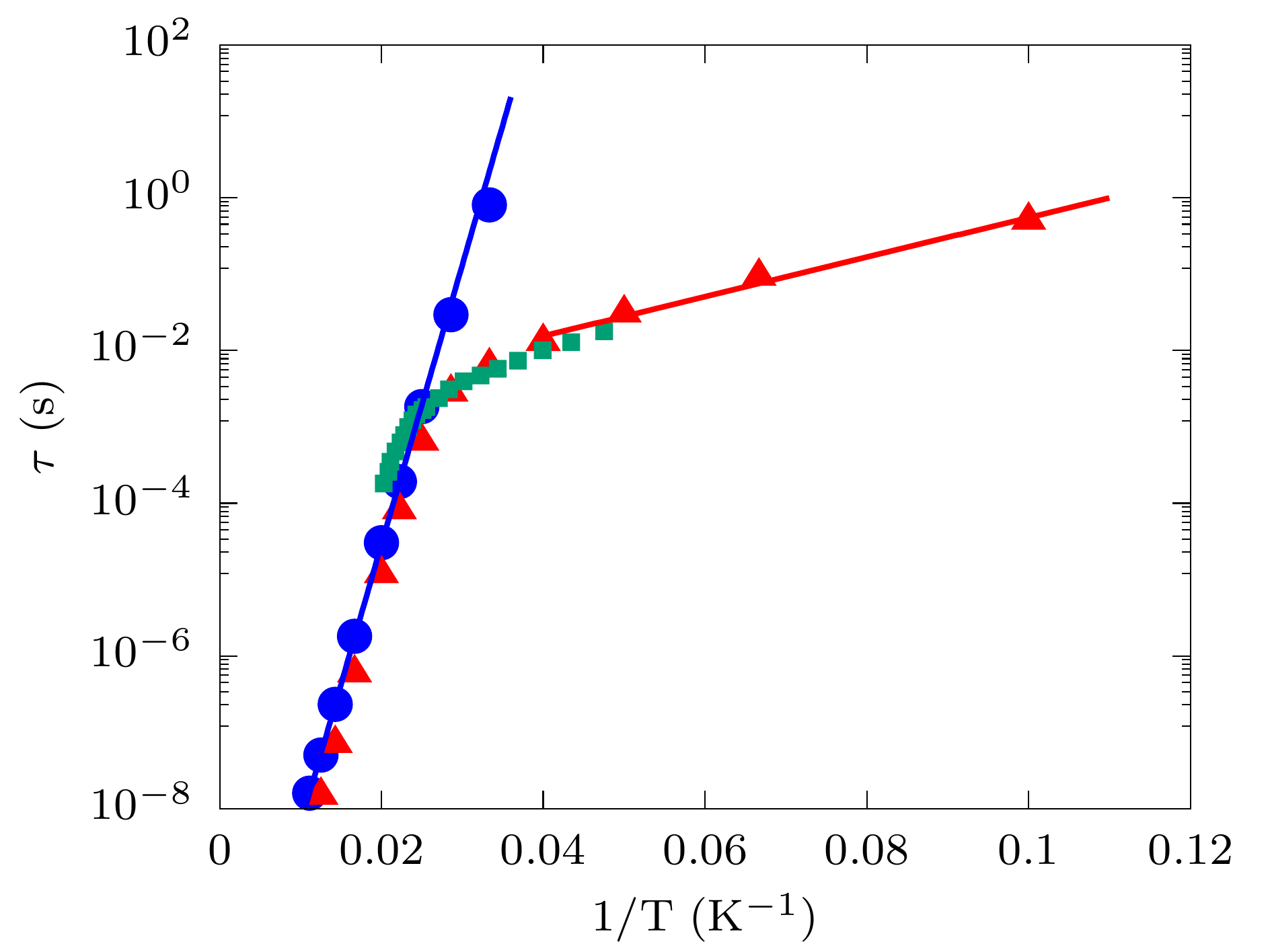}
    \caption{ \textbf{Orbach and Raman contributions to relaxation time.} Comparison of experiment (Green square) with the simulated Orbach (blue circle) and Raman (red triangle up) relaxation for \textbf{9}. Blue solid line represents the fitting of Orbach simulation data with the equation: $\tau_{Orbach}=\tau_0 \mathrm{exp}(U_{eff}/k_BT)$. Similarly, solid red line represents the fitting of Raman simulation data with the equation: $\tau_{Raman}=\tau'_0 \mathrm{exp}(W_{eff}/k_BT)$.}
    \label{image3}
    \hfill
\end{figure}

Now that we have validated our simulations against experimental results, we are in the position to exploit the full power of \textit{ab initio} spin dynamics to disentangle all the contributions to relaxation time. Fig. \ref{image3} shows the decomposition of $\tau$ in terms of Orbach and Raman contributions for compound \textbf{9}, as an illustrative example. The results for all the twelve molecules are reported in Fig S1. In agreement with previous simulations\cite{lunghi2020multiple,briganti2021complete,lunghi2022toward}, Orbach relaxation is found to dominate at high temperature, while Raman mechanism only becomes relevant at low temperature, where phonons in resonance with high-energy spin transitions become too unpopulated. Most importantly, simulations reveal that experimental results for all the molecules were obtained in a regime strongly influenced by Raman relaxation. The only exceptions are represented by \textbf{8} and \textbf{9}, where both magnetization decay experiments and AC magnetometry were employed to sample both the particularly long Orbach relaxation times and the Raman relaxation ones. Surprisingly, even in the case of \textbf{11}, Raman relaxation is found to be the dominant relaxation mechanism despite the first excited KD is low in energy and amenable to promoting Orbach relaxation at low temperature. \textbf{5} is the only system where the Orbach relaxation mechanism is found to be dominating in the experimentally accessible temperature range. In most cases, experimental relaxation times measured with AC magnetometry falls at the transition stage between Orbach and Raman-lead regimes, calling for a reinterpretation of the values of the extracted $U_{eff}$ and $\tau_0$ and a new strategies for determining them in experiments. \textit{Ab initio} simulations offer such an opportunity.

\begin{figure}[!h]
    \centering
    \includegraphics[scale=1]{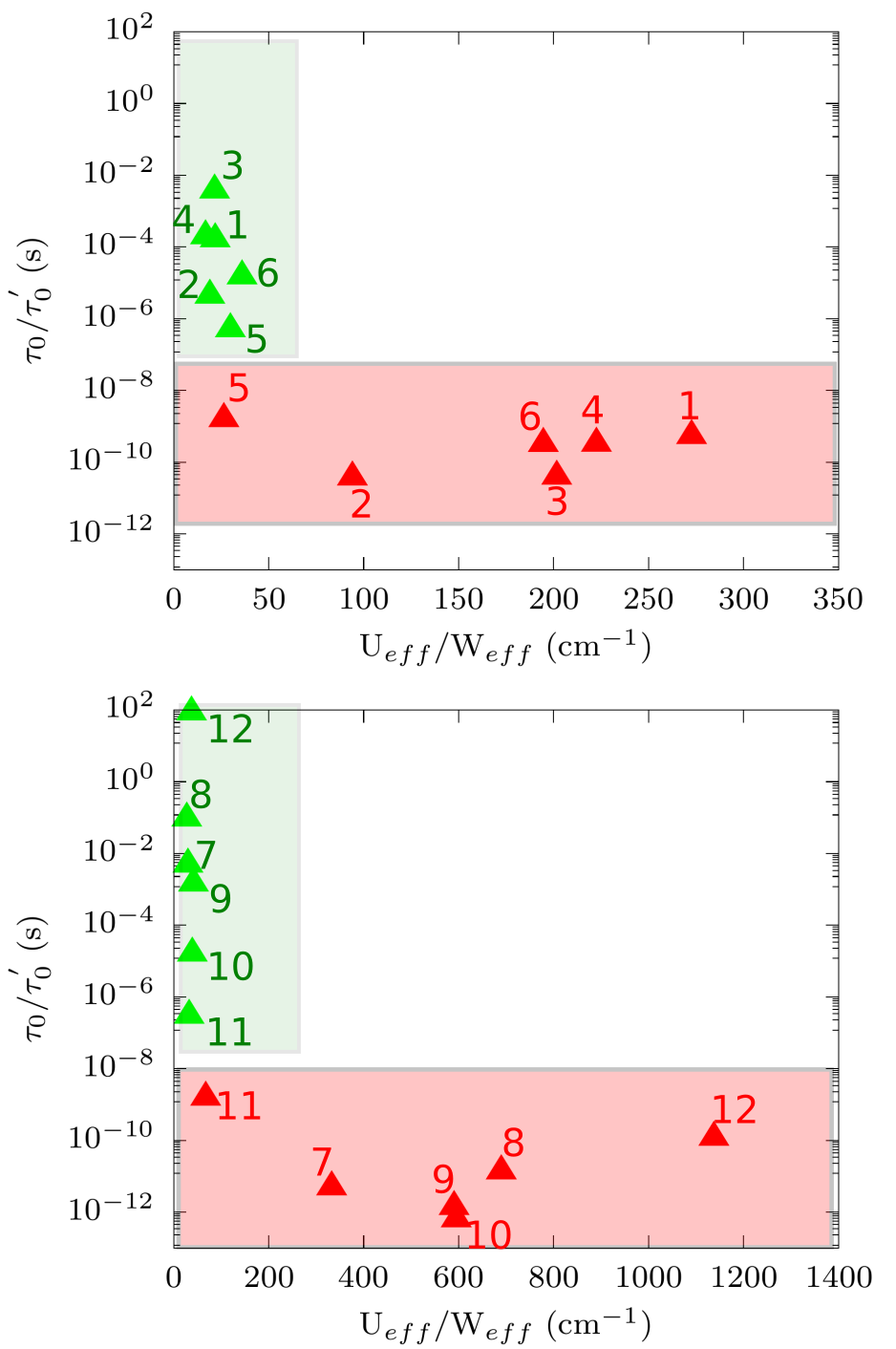}
    \caption{\textbf{Interpretation of spin relaxation with Arrhenius laws.} The figure reports $W_{eff}$ as a function of $\tau^{'}_0$ and $U_{eff}$ as a function of $\tau_0$ for the complexes \textbf{1--6} (top panel), and for the complexes \textbf{7--12} (bottom panel). Green and red triangles represent the simulated values for Orbach and Raman values, respectively. The selected compounds \textbf{1--12} are labelled with the corresponding index number. }
    \label{image4}
    \hfill
\end{figure}

The effective reversal barrier and the pre-exponential factor, $U_{eff}$ and and $\tau_0$, have been extracted from the fitting of the simulated Orbach data with the Arrhenius expression of Eq. \ref{Arr}, as depicted in Fig. \ref{image3}. The results obtained for Raman relaxation follows a more complex mathematical law. Recent literature\cite{lunghi2020multiple,gu2020origins,briganti2021complete} has shown that Raman relaxation is supposed to follow the temperature law
\begin{equation}
    \tau^{-1}_{Raman} = \sum_{i} (\tau'_{0,i})^{-1} ~\frac{e^{W_{eff,i}/k_BT}}{\left(e^{W_{eff,i}/k_BT} -1\right)^2} \:,
    \label{Raman-tau}
\end{equation}
where $W_{eff,i}$ corresponds to the energy of the $i$-th pair of degenerate phonons absorbed and emitted. If only a single pair of phonons contributes to relaxation, the Raman relaxation time also exhibits an Arrhenius behaviour at low temperature ($W_{eff,i} \gg k_BT$),
\begin{equation}
\tau_{Raman}=\tau'_0 e^{W_{eff}/k_BT}\:.
\end{equation}
This behaviour is observed in simulations to a good degree, and we therefore attempt to fit the values of $\tau'_{0}$ and $W_{eff}$ from low-$T$ simulated data, as depicted in Fig. \ref{image3}.

Fig. \ref{image4} reports the values of $\tau_0$, $\tau'_0$, $U_{eff}$ and $W_{eff}$, for all twelve compounds. The $U_{eff}$ for the Co$^{2+}$ systems (\textbf{1--6}) match the energy of the excited KD with $M_{s} = \pm 1/2$. This is in agreement with the fact that the latter is the only available excited KD able to mediate Orbach relaxation. The same is not true for Dy$^{3+}$ complexes, where more than one excited KD is available. Orbach relaxation for the molecules \textbf{7}, \textbf{8}, \textbf{9}, and \textbf{11} is found to be mediated by transitions between ground state ($M_j= \pm 15/2$) and the second excited KD, while for \textbf{10} and \textbf{12}, Orbach transitions occur through the third and fifth excited KD, respectively. Importantly, the values of $\tau_0$ are found to lie in the range $10^{-8}$ - $10^{-12}$ s, a much smaller time window with respect to what extracted from literature (see Fig. \ref{image1}). Raman relaxation provides an opposite picture. $W_{eff}$ is found to span a quite narrow range of small values in the order of tens of cm$^{-1}$. On the other hand, the values of $\tau'_0$ span a range of $\sim$5-6 and $\sim$10 orders of magnitude for Co$^{2+}$ and Dy$^{3+}$, respectively. The comparison between Figs. \ref{image1} and \ref{image4} makes it clear that the experimental values of $U_{eff}$ and $\tau_0$ extracted from literature do not describe the sole Orbach relaxation mechanism but are deeply affected by Raman relaxation. This is particularly important for molecules with a reported small value of $U_{eff}$.

The values of $U_{eff}$ compare nicely with the energy of the excited KDs and present no mystery, but more insights on the other quantities are necessary in order to understand what regulates them. Let us begin from $\tau_0$. According to Eq. \ref{Orbach}, $\tau^{-1}_{0}$ receives contributions from three factors: i) the density of phonons in resonance with the relevant spin transition, ii) their coupling with the magnetic moment, as measured by the derivatives of the effective Hamiltonian coefficients (see Eqs. \ref{sHam} and \ref{CFHam}), and iii) the nature of the static effective Hamiltonian used to compute the matrix elements of the spin-phonon coupling operator. Contributions i) and ii) can be easily assessed by computing the spin-phonon coupling density\cite{lunghi2017intra}, \textit{i.e.} the average coupling of magnetic moment to phonons with a certain energy $\hbar\omega_{\alpha}$. This quantity is defined in ESI and reported in Fig.~S4-S5 for all the compounds. Despite the presence of different features among the twelve compounds, the average values of spin-phonon coupling intensity are not dramatically different. This suggest that contribution iii) is in facts largely responsible for the variance of $\tau_0$. 

\begin{figure}[!htbp]
    \centering
    \includegraphics[scale=1]{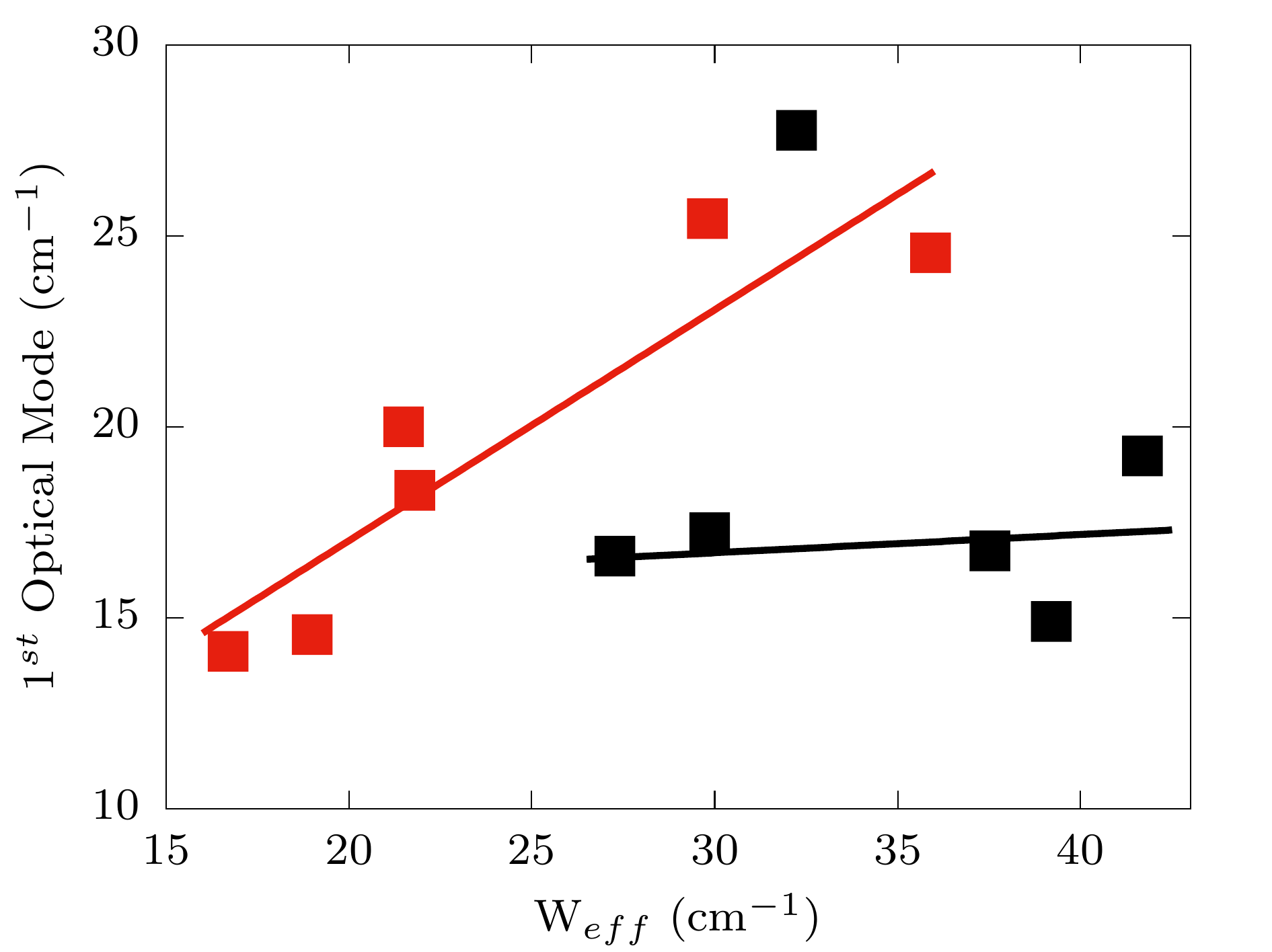}
    \caption{\textbf{Correlation between phonon energies and Raman Arrhenius activation energy.} $W_{eff}$ is plotted as a function of frequency of the first optical mode at the $\Gamma$-point for complexes \textbf{1--6} (red squares) and  \textbf{7--12} (black squares). The correlation coefficient for the two quantities is 0.91 and -0.14 for Co$^{2+}$ and Dy$^{3+}$, respectively. } 
    \label{image5}
    \hfill
\end{figure}

\begin{figure}[!htbp]
    \centering
    \includegraphics[scale=1]{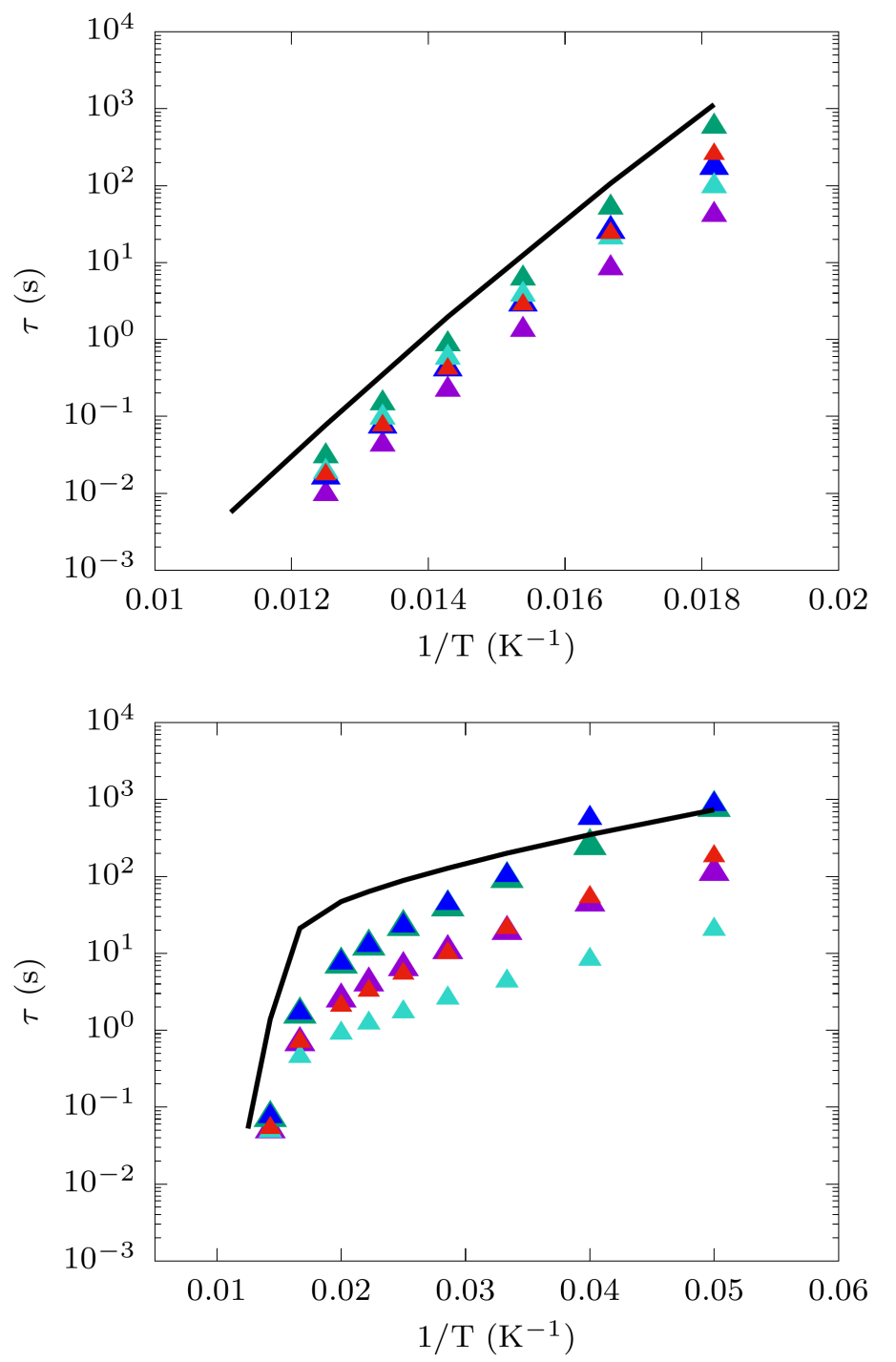}
    \caption{ \textbf{Orbach and Raman relaxation of compound 12 with artificial phonons and spin-phonon coupling coefficients.} The black continuous line represents the original $\tau$ vs $1/T$ for \textbf{12}. Symbols correspond to the simulated values of $\tau$ obtained using the static effective Hamiltonian of \textbf{12} with phonons and spin-phonon coupling of other molecules: 1 (violet), 2 (green), 3 (blue), 4 (turquoise), 5 (red), and 6 (black). Color code for right panel: 7 (violet), 8 (green), 9 (blue), 10 (turquoise), 11 (red), and 12 (black).}
    \label{image6}
    \hfill
\end{figure}

Let us now turn to the analysis of Raman relaxation rates.
According to recent literature, the value of $W_{eff}$ should coincide with the lowest-energy phonons significantly coupled to the magnetic moment\cite{lunghi2020multiple,gu2020origins,briganti2021complete}. In our approximation this generally corresponds to some of the first available optical phonons. Fig. \ref{image5} shows the correlation between the first mode at the $\Gamma$-point and the fitted values of $W_{eff}$. In the case of Co$^{2+}$ a good correspondence between the two quantities is found, validating previous results and further suggesting the importance of low-energy optical vibrations. Surprisingly, the same degree of correlation is not observed for Dy$^{3+}$ compounds, where $W_{eff}$ is found to span slightly larger values of energy that the first optical mode one. Moreover, the values extracted for $W_{eff}$ do not clearly corresponds to peaks in the spin-phonon coupling density. For Dy$^{3+}$ compounds, several excited KDs are at play and it is hard to find a simple rationale to this behaviour. We advance the hypothesis that the presence of very high-energy KDs involved in the virtual state of Dy$^{3+}$ compounds promotes the effect of phonons with higher energy than the first ones available at the $\Gamma$-point. It is still important to remark that the values of $W_{eff}$ are still in the order of tens of cm$^{-1}$, a value commensurate with low-energy optical vibrations and much lower than the energy of excited KDs. Further analysis shows that none of the phonons with energy higher than $\sim$100 - 150 cm$^{-1}$  contribute to spin relaxation (see Figs.~S7) This is in agreement with previous observations that the main factor in the determination of most important phonons is the Bose-Einstein population\cite{lunghi2020multiple,briganti2021complete}. The latter decreases exponentially as the energy of vibrations increases, thus leaving the lowest-energy available modes to drive spin relaxation. 

From a qualitative point of view, Raman relaxation rate depends on similar quantities to the Orbach one, and $\tau'_{0}$ is influenced by both the spin-phonon coupling density and the nature of static effective Hamiltonian. As discussed for Orbach relaxation, the differences in the former quantity across the twelve molecules are not dramatic and cannot account for the large span of values predicted for $\tau'_{0}$. The variance in values of $\tau'_0$ must therefore come i) the nature of the eigenstates of the static effective Hamiltonian used to compute the matrix elements of the spin-phonon coupling operator, and ii) the energy of the excited KDs appearing at the denominator of Eq. \ref{Raman}. In both cases, these quantities are intimately linked to the nature of the static crystal field and the intensity of the zero-field splitting. \\

According to this analysis, the static effective Hamiltonian stands out as the most important contribution to both $\tau_{0}$ and $\tau'_{0}$. In order to provide a conclusive proof of this claim we perform a simulation of $\tau$ for Dy$^{3+}$ compounds where the static effective Crystal field of \textbf{12} is used together with the phonons and spin-phonon coupling of the other compounds. Results for Orbach and Raman relaxation in this artificial conditions are reported in the top and bottom panels of Fig. \ref{image6}, respectively. Orbach rates are found to all fall within one order of magnitude, with the genuine value of \textbf{12} as the slowest one. In the case of Raman relaxation a similar behaviour is observed, except for a larger variance of relaxation times, which now span up to two orders of magnitude. The latter value should be compared with a total variation of $\sim$6 orders of magnitude in normal conditions (see for instance the values of $\tau$ at 20 K in the right panel of Fig. \ref{image2}). We attribute this larger sensitivity to the details of spin-phonon coupling and the vibrational density of state in Raman relaxation to the fact that these quantities appear to a power law of four instead of two as in Orbach relaxation. Performing the same analysis for the Co$^{2+}$ SMMs we find the same qualitative behaviour but a larger span of relaxation times, reaching almost two orders of magnitude for Orbach rates and four orders or magnitude for Raman relaxation (see Fig. S3). This analysis conclusively demonstrates that the static effective Hamiltonian is the most important contribution to $\tau$ across all relaxation regimes. However, the details of spin-phonon coupling add on top of that to fine-tune the value of $\tau$ and in the case of Raman relaxation significantly contribute to determining the magnetic moment lifetime.

\section*{Discussion and Conclusions}

Molecular magnetic anisotropy has been identified as a key ingredient for slow spin relaxation since the very first observation of magnetic hysteresis in a mixed-valence Mn$_{12}$ cluster, now 30 years ago\cite{sessoli1993magnetic}. Decades of success stories of molecular magnetism have marked the synthesis of compounds with previously unimaginably large zero-field splitting values, reaching record values of above 2000 K\cite{zabala2021single,guo2018magnetic}. These results effectively translated into the possibility of stabilizing the molecular magnetic moment over times scales of hundreds of seconds at temperatures as high as 80 K\cite{guo2018magnetic}. Despite the past success, the field of single-molecule magnets now find itself at a critical stage. It has been argued that the strategies that have led to large $U_{eff}$ in single-ion complexes cannot realistically lead to much further improvement and that new approaches to increasing $\tau$ must be found\cite{reta2021ab,briganti2021complete}. We argue that further progress in the design of molecular compounds with long spin lifetime is to be found in a better understanding of the entire process of spin relaxation. In this contribution we have highlighted how the unique focus on $U_{eff}$ has led to overlooking many important aspects of spin relaxation. The analysis of the pre-exponential factor $\tau_0$ is a striking example of such a situation, where years of information available in literature have remained unexplored.

In order to move out from this impasse, new tools are needed. In this work we built on our recent contributions and have further shown how \textit{ab initio} spin dynamics simulations provide an effective way to obtain unprecedented quantitative details on the spin relaxation process of single-molecules magnets. Here we have fully disentangled the various contributions to both Orbach and Raman relaxation posing an end to years of debate on the subject. We have demonstrated that differently from the predictions of the canonical theory of spin-phonon relaxation\cite{gatteschi2006molecular} and from what arise from literature, there is no simple correlation between $U_{eff}$ and $\tau_0$. Moreover, simulations made it possible to individuate realistic ranges for both these quantities and those relative to Raman relaxation, when interpreted as an Arrhenius process. We anticipate that this information will play a fundamental role in the interpretation of future experiments and will provide a guide to the assignment of Orbach and Raman relaxation mechanisms.

The unprecedented effort of studying twelve crystals of SMMs made it possible to take a first glimpse to the correlation between chemical structure and spin relaxation. Despite the large differences between the selected molecular compounds, we demonstrated that Orbach spin relaxation rate is mostly determined by the zero-field splitting. Now that spin relaxation can be correctly interpreted, it becomes clear that the natural variations of $\tau_0$ are too small to overcome the effect of $U_{eff}$. The little dependence of the Orbach rate over the details of spin-phonon coupling suggests that little improvement can be achieved by serendipitously tuning vibrational modes. As highlighted in Fig.~S2, the vibrational density of states below $\sim$ 1500-1700 cm$^{-1}$ is densely populated and only a very delicate tailoring can bring phonons and spin transitions completely out of resonance\cite{ullah2019silico}. Similarly, the static zero-field splitting is found to also strongly affect Raman relaxation rate. However, in this case the features of spin-phonon coupling and low-energy vibrations significantly contribute to modulating relaxation time and may offer a way forward to further improvement. These findings thus shift the attention to Raman relaxation as the most feasible improvement that can be obtained in single-ion SMMs, and point to the necessity of better understanding how the low-energy vibrational structure of coordination compounds can be chemically engineered. 
We envision that a tighter synergy between experimental techniques and simulations hold the key to significant advances in this direction. For instance, inelastic neutron scattering\cite{garlatti2020unveiling} and terahertz spectroscopy\cite{atzori2017spin,albino2019first,albino2021temperature,de2021exploring,ma2022ligand} have already been used to provide unique insights on the low energy part of the vibrational spectrum. Similarly, far-infrared magneto spectroscopy can be used to provide insights on the spin-phonon coupling of vibrations in close resonance to the spin transitions responsible for Orbach relaxation\cite{rechkemmer2016four,moseley2018spin,blockmon2021spectroscopic,kragskow2022analysis}. Once these techniques are combined with \textit{ab initio} simulations, a clear picture of how vibrations couple with spin is possible.

Last but not least, we would like to stress out that the present findings were made possible by studying a large number of complexes on the same footing. We envision that overcoming the traditional approach of studying a single molecule at the time or homologous small series of compounds can lead to a much better understanding of structure-properties relations across the chemical space. Moreover, we have here shown that accurate predictions down to one order of magnitude of $\tau$ are now possible for a large breadth of chemical compositions and zero-field splittings. This level of accuracy is already enough for enabling a blind exploration of the chemical space in search of new compounds. We envision that the further combination of \textit{ab initio} spin dynamics with machine learning\cite{janet2017predicting,lunghi2020limit,lunghi2020multiple,lunghi2020surfing,nguyen2022predicting} and high-throughput strategies\cite{nandy2021computational,duan2021data} may lead to a significant reduction in its computational cost, potentially leading to a paradigm shift in the way we design molecular compounds.

In conclusion, we have provided a full \textit{ab initio} description of spin relaxation in twelve single-molecule magnets based on Co$^{2+}$ and Dy$^{3+}$ ions that represent the extent of the chemical space explored so far in this field. Our simulations made it possible to resolve a conflicting interpretation of experimental results in terms of Orbach and Raman mechanism and to rationalize all the main contributions to relaxation. We found that zero-field splitting is the main figure of merit determining spin relaxation, but the efficiency of Raman relaxation is also significantly influenced by the details of the low-energy vibrational spectrum. We anticipate that these results will significantly inform future synthetic strategies. \\

\vspace{0.2cm}
\noindent
\textbf{Acknowledgements}\\
This project has received funding from the European Research Council (ERC) under the European Union’s Horizon 2020 research and innovation programme (grant agreement No. [948493]). Computational resources were provided by the Trinity College Research IT and the Irish Centre for High-End Computing (ICHEC). We acknowledge Matteo Briganti for the useful discussions.


\end{document}


\date{}

\title{\textbf{Supporting Information-}\\\textbf{Unravelling the contributions to spin-lattice relaxation in Kramers single-molecule magnets}}

\author{Sourav Mondal} 
\author{Alessandro Lunghi \thanks{lunghia@tcd.ie}}
\affil{School of Physics, AMBER and CRANN Institute, Trinity College, Dublin 2, Ireland}

\maketitle


\begin{table}
    \centering
    \caption{Table S1: \textbf{Cobalt SMM compounds}. The table reports $U_{eff}$ and $\tau$ for the 56 Cobalt SMMs individuated in literature.}
    \begin{tabular}{|c c c| c c c |}
    \hline
    $U_{eff}$ (cm$^{-1}$)  & $\tau$ (s) & REF & $U_{eff}$ (cm$^{-1}$)  & $\tau$ (s) & REF \\
    \hline
56.300000 &   6.0000e-10 & \cite{huang2014field}      &           25.920000 &   9.4000e-11 &  \cite{boca2014simple}     \\
20.700000 &   1.2000e-06 & \cite{huang2014field}      &           39.400000 &   1.3000e-08 &  \cite{cao2013mononuclear}     \\
62.300000 &   8.7000e-11 &  \cite{huang2014field}     &           21.270000 &   4.6500e-10 &  \cite{saber2014ligands}     \\
11.100000 &   3.6000e-06 &   \cite{jurca2011single}   &           22.670000 &   1.5000e-08 &   \cite{saber2014ligands}      \\
16.700000 &   5.1000e-07 &   \cite{jurca2011single}   &           27.800000 &   5.9800e-11  & \cite{boca2014simple} \\
23.100000 &   2.5000e-07 & \cite{palacios2017analysis}     &           25.800000 &   1.2000e-09 & \cite{yang2013inspiration}  \\
24.100000 &   2.3000e-07 &  \cite{palacios2017analysis}    &           24.300000 &   2.1000e-10 & \cite{yang2013inspiration}  \\
2.800000 &   7.4000e-02 &  \cite{habib2013influence}     &            20.800000 &   6.0000e-09 & \cite{yang2013inspiration}  \\
11.800000 &   5.9000e-06 & \cite{habib2013influence}     &           21.100000 &   7.0000e-10 & \cite{zadrozny2013slow}  \\
2.100000 &   1.0000e-01 & \cite{habib2013influence}      &           317.000000 &   4.6000e-11 & \cite{yao2017two}  \\ 
23.000000 &   4.0000e-06 & \cite{gomez2013mononuclear}  &        21.100000 &   1.0000e-06 & \cite{zadrozny2013slow}  \\
8.700000 &   8.0000e-06 &   \cite{gomez2013mononuclear} &         19.100000 &   3.0000e-06 & \cite{zadrozny2013slow}  \\
16.200000 &   4.0000e-07 &  \cite{vallejo2012field}      &          33.900000 &   4.5000e-06 & \cite{fataftah2014mononuclear}  \\
24.000000 &   2.0000e-10 &    \cite{zadrozny2011slow}     &          75.800000 &   1.0000e-07 & \cite{zhu2013zero}      \\
24.000000 &   1.9000e-09 &  \cite{zadrozny2012slow}      &          7.900000 &   6.1000e-06 & \cite{peng2017field}      \\
22.900000 &   3.7000e-10 & \cite{huang2013field} &        14.500000 &   1.0000e-06 &   \cite{echr2014slow}    \\
10.900000 &   8.9000e-07 & \cite{palacios2020tuning} &       230.000000 &   7.6000e-11 & \cite{rechkemmer2016four}  \\
59.900000 &   1.4000e-09 & \cite{herchel2014slow}  &        191.000000 &   8.8700e-10 & \cite{pavlov2016polymorphism}  \\
17.000000 &   1.5000e-06 &   \cite{chen2014slow}      &          122.000000 &   2.6500e-09 & \cite{pavlov2016polymorphism}  \\
308.000000 &   8.9000e-10 & \cite{yao2017two} &         29.200000 &   1.4000e-07 &   \cite{rigamonti2018pseudo}    \\
 413.000000 &   1.1000e-10 & \cite{yao2017two}  &   16.400000 &   4.8000e-06 & \cite{mondal2018influence}  \\
 43.000000 &   8.4000e-10 & \cite{ziegenbalg2016cobalt}   &    21.000000 &   4.6000e-08 & \cite{shao2017structural}      \\
 29.800000 &   1.8000e-10 & \cite{nemec2017magnetic}      &       21.000000 &   1.0000e-07 &   \cite{nemec2017magnetic}    \\
 13.600000 &   5.8000e-05 & \cite{mondal2017probing}  &    20.000000 &   2.0000e-09 & \cite{shao2017structural}      \\
 5.700000 &   4.6000e-05 & \cite{mondal2017probing}   &     48.000000 &   1.4000e-15 &  \cite{rajnak2018octahedral}     \\
 23.300000 &   7.4000e-06 & \cite{mondal2018influence}  &    450.000000 &   1.7900e-09 & \cite{bunting2018linear} \\
 19.700000 &   5.6000e-06 & \cite{mondal2018influence}   &    87.000000 &   1.1000e-09 & \cite{chakarawet2018large}      \\
 10.400000 &   5.6000e-06 & \cite{zhou2018slow}  &   36.000000 &   5.6000e-10 & \cite{ziegenbalg2016cobalt}   \\      
\hline
     \end{tabular}
    \label{tab:my_label}
\end{table}

\begin{table}
    \centering
    \caption*{TABLE S2: \textbf{Crystal Field Parameters} Crystal field parameter with rank of \textit{l} = 2, 4, and 6 are reported for \textbf{7}, \textbf{8}, and \textbf{9}.}
     \begin{tabular}{|c c |c|c|c|}
     \hline
     \textit{l} & \textit{m} & \textbf{7} & \textbf{8} & \textbf{9} \\
     \hline
        2 &  -2 &      0.0116382160 &     -0.0826793069 &     0.0549505597  \\
     2 &  -1 &      3.4257015712 &     -0.1527753972 &    -0.0838634963  \\
     2 &   0 &     -5.8586792439 &     -8.7796677967 &    -7.6110790511  \\
     2 &   1 &     -0.0099168347 &      6.1522171892 &     5.8584518358  \\
     2 &   2 &      0.9673412592 &     -0.5972375274 &     0.5796986733  \\
     \hline
     4 &  -4 &      0.0027628929 &     -0.0006242182 &    -0.0000287162  \\
     4 &  -3 &     -0.0106387059 &     -0.0015804297 &     0.0000880570  \\
     4 &  -2 &      0.0048945377 &      0.0013029424 &     0.0009347407  \\
     4 &  -1 &      0.1072086362 &      0.0001074779 &    -0.0003398877  \\
     4 &   0 &     -0.1846407837 &     -0.0445629700 &    -0.0376264303  \\
     4 &   1 &      0.0001298547 &      0.1522592206 &     0.1440703047  \\
     4 &   2 &     -0.0011175341 &     -0.0810106000 &    -0.0856459975  \\
     4 &   3 &     -0.0062517828 &     -0.0047288351 &    -0.0102090366  \\
     4 &   4 &      0.0089904206 &     -0.0006351176 &    -0.0101781822  \\
     \hline
     6 &  -6 &     -0.0004116294 &      0.0000051326 &    -0.0000052892  \\
     6 &  -5 &     -0.0001129912 &      0.0000270237 &     0.0000054361  \\
     6 &  -4 &      0.0003933766 &      0.0000084263 &    -0.0000062021  \\
     6 &  -3 &      0.0001579314 &      0.0000078332 &     0.0000089929  \\
     6 &  -2 &      0.0000512990 &     -0.0000113149 &     0.0000015686  \\
     6 &  -1 &     -0.0019701136 &     -0.0000029264 &    -0.0000011981  \\
     6 &   0 &      0.0011167029 &     -0.0007246044 &    -0.0007860787  \\
     6 &   1 &     -0.0000500032 &     -0.0004747567 &    -0.0003637745  \\
     6 &   2 &     -0.0005854917 &      0.0000051248 &    -0.0002102721  \\
     6 &   3 &      0.0001137287 &      0.0003188668 &     0.0004103597  \\
     6 &   4 &     -0.0006425285 &      0.0005579449 &     0.0005183958  \\
     6 &   5 &     -0.0000810403 &      0.0006447734 &     0.0007425981  \\
     6 &   6 &      0.0005401116 &      0.0010857455 &     0.0010163157  \\

       \hline
    \end{tabular}
    \label{T1}
\end{table}

\begin{table}
    \centering
    \caption*{TABLE S3: \textbf{Crystal Field Parameters} Crystal field parameter with rank of \textit{l} = 2, 4, and 6 are reported for \textbf{10}, \textbf{11}, and \textbf{12}.}
     \begin{tabular}{|c c |c|c|c|}
     \hline
     \textit{l} & \textit{m} & \textbf{10} & \textbf{11} & \textbf{12} \\
     \hline
     2 &  -2 &       -2.3065713232 &     0.3577254846 &     11.0687564278 \\
     2 &  -1 &       -5.3072539097 &     0.2173623886 &      0.4363723381 \\
     2 &   0 &       -1.9292231809 &    -7.8353612716 &     11.1144759237 \\
     2 &   1 &       -9.8318549040 &     0.1053194000 &     -0.8325478201 \\
     2 &   2 &       -0.7329379913 &    -0.1675615285 &    -13.5092124727 \\
     \hline
     4 &  -4 &        0.0628592935 &     0.0013903411 &      0.0369873249 \\
     4 &  -3 &       -0.0764898133 &     0.0071693547 &      0.0005814264 \\
     4 &  -2 &       -0.0063222450 &     0.0010805203 &      0.0006340423 \\
     4 &  -1 &        0.0452744858 &    -0.0001020090 &     -0.0001981386 \\
     4 &   0 &       -0.0323086070 &     0.1078582823 &     -0.0156835041 \\
     4 &   1 &       -0.0756484466 &    -0.0002993690 &      0.0017361378 \\
     4 &   2 &       -0.0004469365 &    -0.0017197848 &      0.0101241854 \\
     4 &   3 &        0.0442063120 &     0.0006781750 &     -0.0010780319 \\
     4 &   4 &        0.0624170051 &     0.0008755701 &     -0.0033934035 \\
     \hline
     6 &  -6 &       -0.0000185657 &     0.0009055887 &      0.0017818310 \\
     6 &  -5 &        0.0004616557 &     0.0000327369 &      0.0000588971 \\
     6 &  -4 &        0.0009230402 &    -0.0000029653 &      0.0001654676 \\
     6 &  -3 &       -0.0000185859 &     0.0001999129 &     -0.0000096911 \\
     6 &  -2 &        0.0002903747 &     0.0000419533 &      0.0001474241 \\
     6 &  -1 &        0.0000174419 &     0.0000304756 &      0.0000178138 \\
     6 &   0 &       -0.0007014251 &    -0.0002546643 &      0.0000674272 \\
     6 &   1 &        0.0000595712 &     0.0000235277 &     -0.0000208904 \\
     6 &   2 &       -0.0004374130 &     0.0000336993 &      0.0001148088 \\
     6 &   3 &       -0.0003279038 &     0.0002739388 &     -0.0000414967 \\
     6 &   4 &        0.0003567190 &    -0.0000416915 &     -0.0000176056 \\
     6 &   5 &        0.0005184329 &    -0.0000170437 &      0.0000504922 \\
     6 &   6 &       -0.0008630159 &    -0.0008856957 &      0.0003649024 \\
        \hline
    \end{tabular}
    \label{T2}
\end{table}


\begin{table}
   \caption*{TABLE S4: \textbf{Dy-SMMs Kramers Doublet Energies.} Computed energy (in cm$^{-1}$) of the KDs for the complexes 7-12}
    \centering
    \begin{tabular}{c c c c c c c c c}
    \hline
         & KD1 & KD2 & KD3 & KD4 & KD5 & KD6 & KD7 & KD8 \\
    \hline 
      \textbf{7}  & 0.00 & 293.84 & 437.94 & 498.52 & 523.02 & 582.70 & 638.90 & 706.85 \\
      \textbf{8} & 0.00 & 394.41 & 636.50 & 741.33 & 768.63 & 783.44 & 812.12 & 864.21 \\
      \textbf{9} & 0.00 & 370.51 & 573.19 & 640.32 & 702.97 & 718.00 & 728.96 & 810.70 \\
      \textbf{10} & 0.00 & 240.94 & 452.52 & 573.31 & 620.83 & 704.76 & 761.92 & 968.38 \\
      \textbf{11} & 0.00 & 10.55 & 68.74 & 166.01 & 274.85 & 393.68 & 472.93 & 520.95 \\
      \textbf{12} & 0.00 & 451.00 & 752.26 & 975.82 & 1159.41 & 1308.57 & 1413.75 &1473.01  \\
      \hline
      \end{tabular}
    \label{T3}
\end{table}

\begin{table}
    \centering
    \caption*{TABLE S5: \textbf{Dy-SMMs g-Tensors.} g-Tensors corresponding to each KD for the complexes 7--12}
    \begin{tabular}{c c c c c c c c}
    \hline
      & & \textbf{7} & \textbf{8} & \textbf{9} & \textbf{10} & \textbf{11} & \textbf{12} \\
      \hline
         & g$_x$ & 0.000018 & 0.000228  & 0.001008 &0.003047 &0.030002 & 0.000002 \\
       KD1  & g$_y$ & 0.000113 & 0.000283 & 0.001364 &0.004857 &0.053263 & 0.000003 \\
          & g$_z$ & 19.861855 & 19.865906 & 19.860118 &19.636431 &19.527463 & 19.579332 \\
          \hline
               & g$_x$ & 0.054184 & 0.052223 & 0.107678 &0.036654 &0.020140 & 0.000011 \\
       KD2  & g$_y$ & 0.057415 & 0.059450 & 0.151784 &0.041708 &0.064976 & 0.000016 \\
          & g$_z$ & 17.054415 & 16.987918 & 16.919610 &16.900190 &18.408833 & 16.774470 \\    
          \hline
            & g$_x$ & 0.661522 & 0.267119 & 2.086339 &0.207555 &0.044689 & 0.001914 \\
       KD3  & g$_y$ & 3.344376 & 0.569134 & 4.869881 &0.255673 &0.082302 & 0.002125 \\
            & g$_z$ & 16.019078 & 13.919003 & 11.726564 &14.588700 &14.549004 & 14.134042 \\    
          \hline
            & g$_x$ & 3.414824 & 5.368526 & 1.830876 &0.988891 &0.253496 & 0.002284 \\
       KD4  & g$_y$ & 3.776073 & 6.427398 & 6.698137 &1.534766 &0.284410 & 0.005746 \\
            & g$_z$ & 7.933602 & 6.983514 & 8.539995 &14.077576 &11.751497 & 11.501928 \\    
            \hline
            & g$_x$ & 0.280516 & 0.842745 & 0.730344 &2.920243 &4.015049 & 0.107421 \\
       KD5  & g$_y$ & 1.170205 & 1.670309 & 1.372234 &5.431001 &4.483045 & 0.112281 \\
            & g$_z$ & 12.557556 & 16.684468 & 12.244360 &9.773901 &8.364658 & 8.900419 \\    
            \hline
            & g$_x$ & 1.240577 & 0.561377 & 0.272030 &1.903553 &3.256554 & 0.174136 \\
       KD6  & g$_y$ & 4.232480 & 4.835739  & 0.324692 &3.874651 &5.255842 & 0.384298 \\
            & g$_z$ & 8.486545 & 11.519193 & 18.375360 &10.662750 &5.679092 & 6.326653 \\    
            \hline
            & g$_x$ & 1.484707 & 0.691294 & 1.295311  &0.938560 &2.734825 & 3.638125 \\
       KD7  & g$_y$ & 2.233398 & 3.276402 &1.893175 &2.582974 &3.604480 & 3.658512 \\
            & g$_z$ & 4.397454 & 9.927467 &12.870477 &15.823137 &4.377979 & 4.033277 \\    
            \hline
            & g$_x$ & 0.520194 & 0.397944 &0.034986 &0.006481 &1.131326 & 1.128157 \\
       KD8  & g$_y$ & 1.252197 & 1.185749 &0.067722 &0.008248 &6.846465 & 6.405268 \\
            & g$_z$ & 16.474694 & 17.046018 &18.600018 &19.733254 &13.982897 & 14.117889 \\    
            \hline           
    \end{tabular}
    \label{T4}
\end{table}

\begin{table}[h!]
    \centering
    \caption*{TABLE S6: \textbf{Dilution data and value of magnetic field used in the Experiment }}
  
\begin{tabular}{|c|c|c|}
   \hline
        System &  Concentration (in $\%$)  & Magnetic Field (Oe) \\ 
        \hline
  \textbf{1}  & 100.0 & 0 \\
  \textbf{2}  & 100.0 & 0  \\
  \textbf{3}  & 100.0 & 0 \\
  \textbf{4}  & 100.0 & 0 \\
  \textbf{5}  & 100.0 & 1000  \\
  \textbf{6} & 100.0 & 0\\
  \textbf{7} & 100.0 & 0 \\
  \textbf{8} & 100.0 & 0 \\
  \textbf{9} & 100.0 & 0 \\
  \textbf{10} & 100.0 & 0\\
  \textbf{11} & 100.0 & 1500 \\
  \textbf{12} & 100.0 & 0\\
        \hline
    \end{tabular}
    \label{T5}
\end{table}

\begin{figure}[t]
    \centering
    \includegraphics[scale=1]{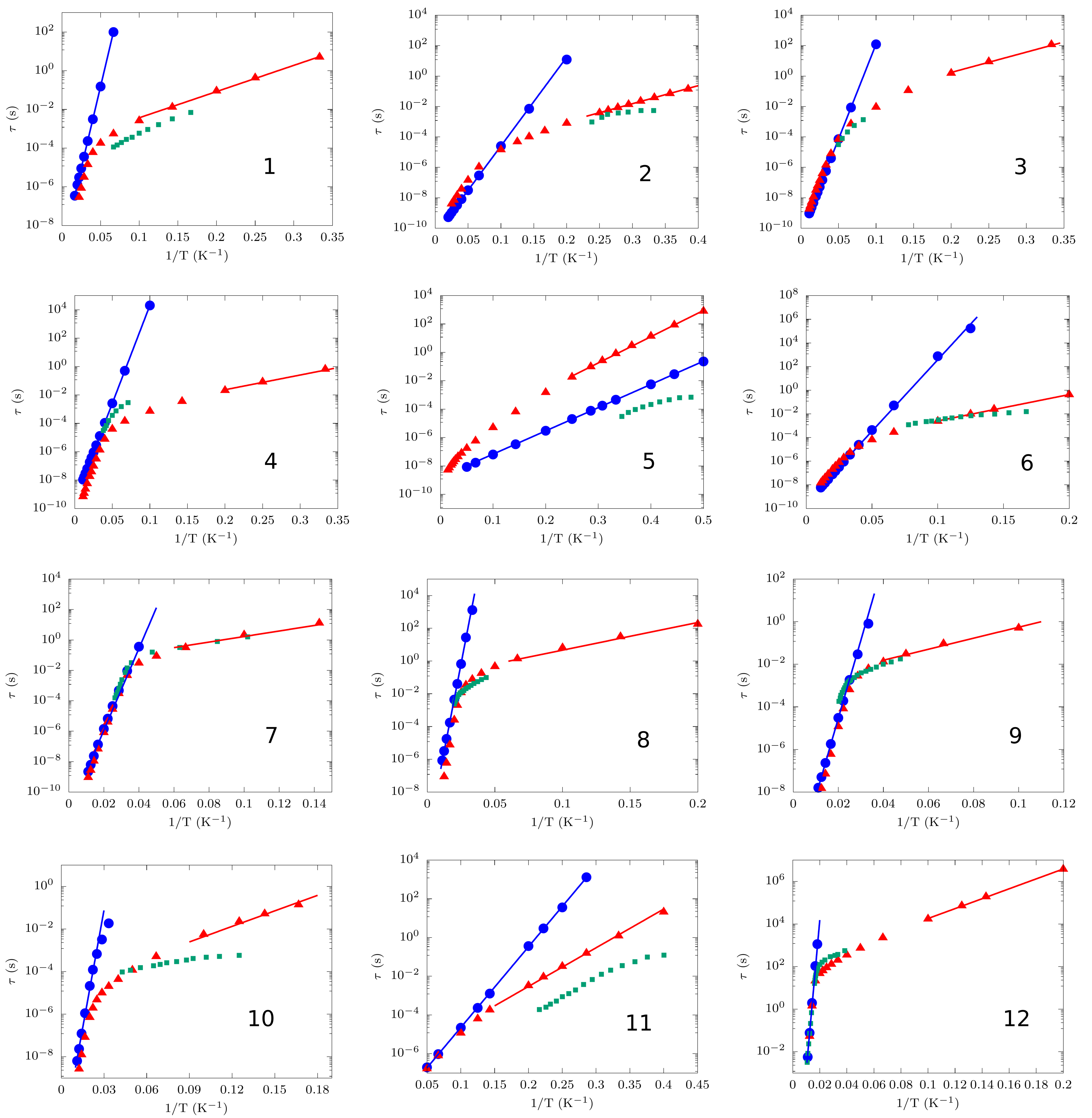}
    \caption*{FIG. S1: \textbf{Orbach and Raman contributions to relaxation time} Comparison of experiment (green square) with the simulated Orbach (blue circle) and Raman (red triangle up) relaxation for \textbf{1-12}. Blue solid line represents the fitting of Orbach simulation data with the equation: $\tau_{Orbach}=\tau_0~\mathrm{exp}(U_{eff}/k_BT)$. Similarly, solid red line represents the fitting of Raman simulation data with the equation: $\tau_{Raman}=\tau^{'}_0~\mathrm{exp}(W_{eff}/k_BT)$. }
    \label{S1}
\end{figure}

\begin{figure}[t]
    \centering
    \includegraphics[scale=1]{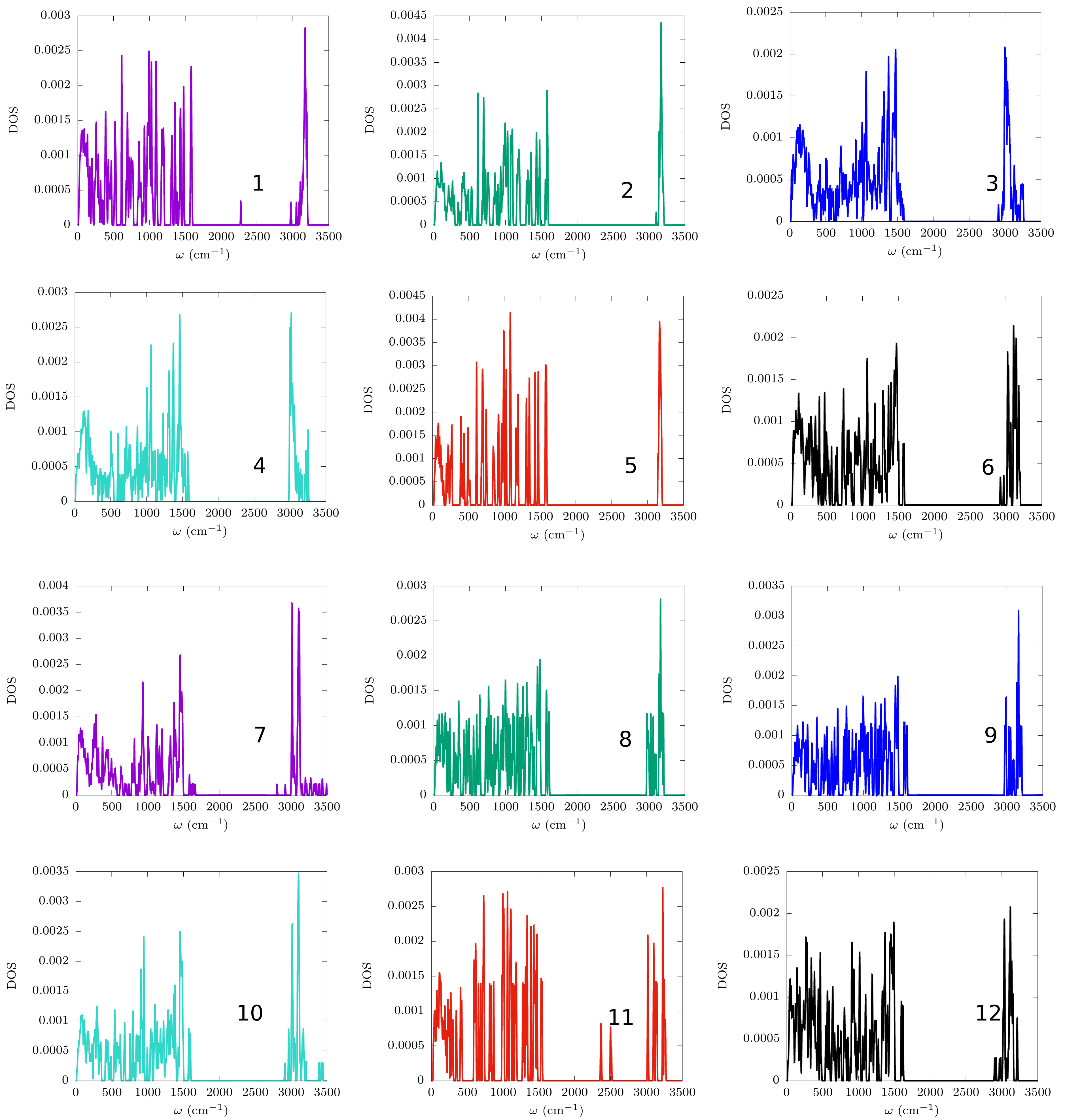}
    \caption*{FIG. S2: \textbf{Phonon density of states for 1--12.} The total phonon density of states (DOS) as a function of the phonon frequency with Gaussian smearing of 5 cm$^{-1}$. Color code for top two panels: violet (\textbf{1}), green (\textbf{2}), blue (\textbf{3}),  turquoise (\textbf{4}), red (\textbf{5}), and black (\textbf{6}). Color code for bottom two panels:  violet (\textbf{7}), green (\textbf{8}), blue (\textbf{9}),  turquoise (\textbf{10}), red (\textbf{11}), and black (\textbf{12}). }
    \label{S2}
\end{figure}

\begin{figure}[t]
    \centering
    \includegraphics[scale=1]{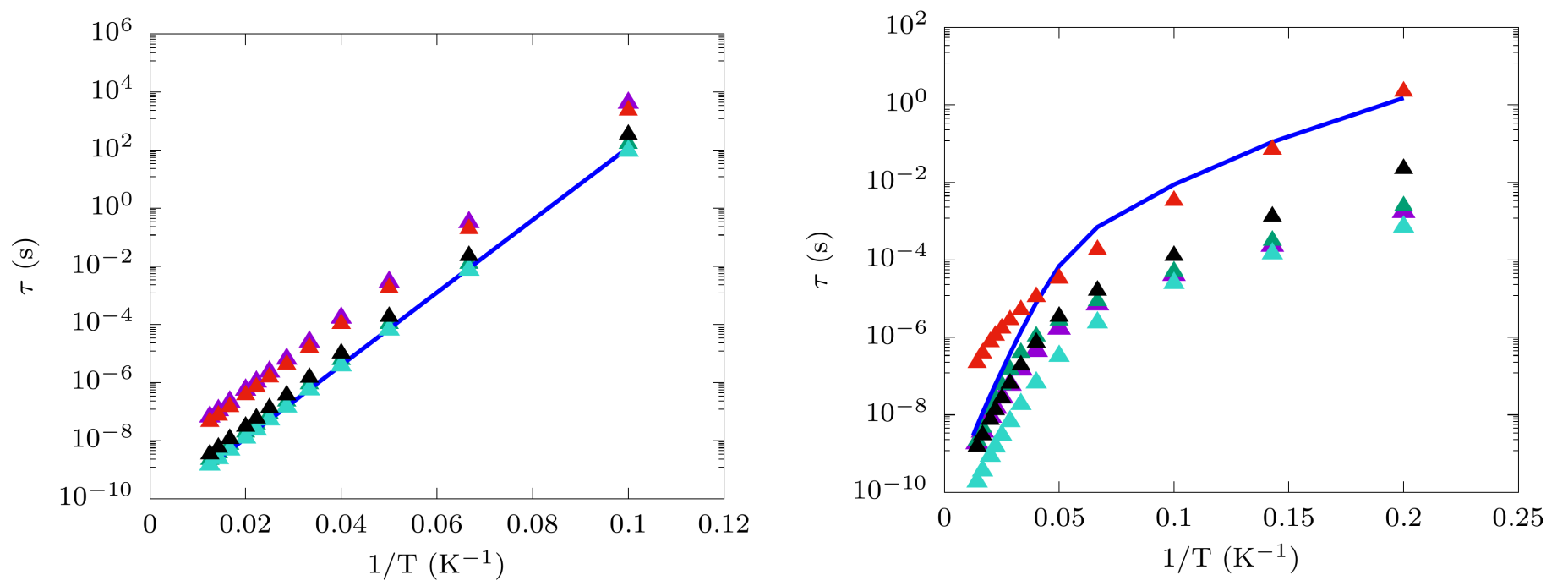}
    \caption*{FIG. S3: \textbf{Relaxation of 3 with artificial phonons and spin-phonon coupling.} The blue continuous line represents the original $\tau$ vs $1/T$ for \textbf{3}. Orbach and Raman relaxation rates are reported in the left and right panel, respectively. Symbols corresponds to the simulated values of $\tau$ obtained using the static effective Hamiltonian of \textbf{3} with phonons and spin-phonon coupling of other molecules: \textbf{1} (violet), \textbf{2} (green) \textbf{4} (turquoise), \textbf{5} (red), and \textbf{6} (black).   }
    \label{S3}
\end{figure}

\begin{figure}[t]
    \centering
    \includegraphics[scale=1]{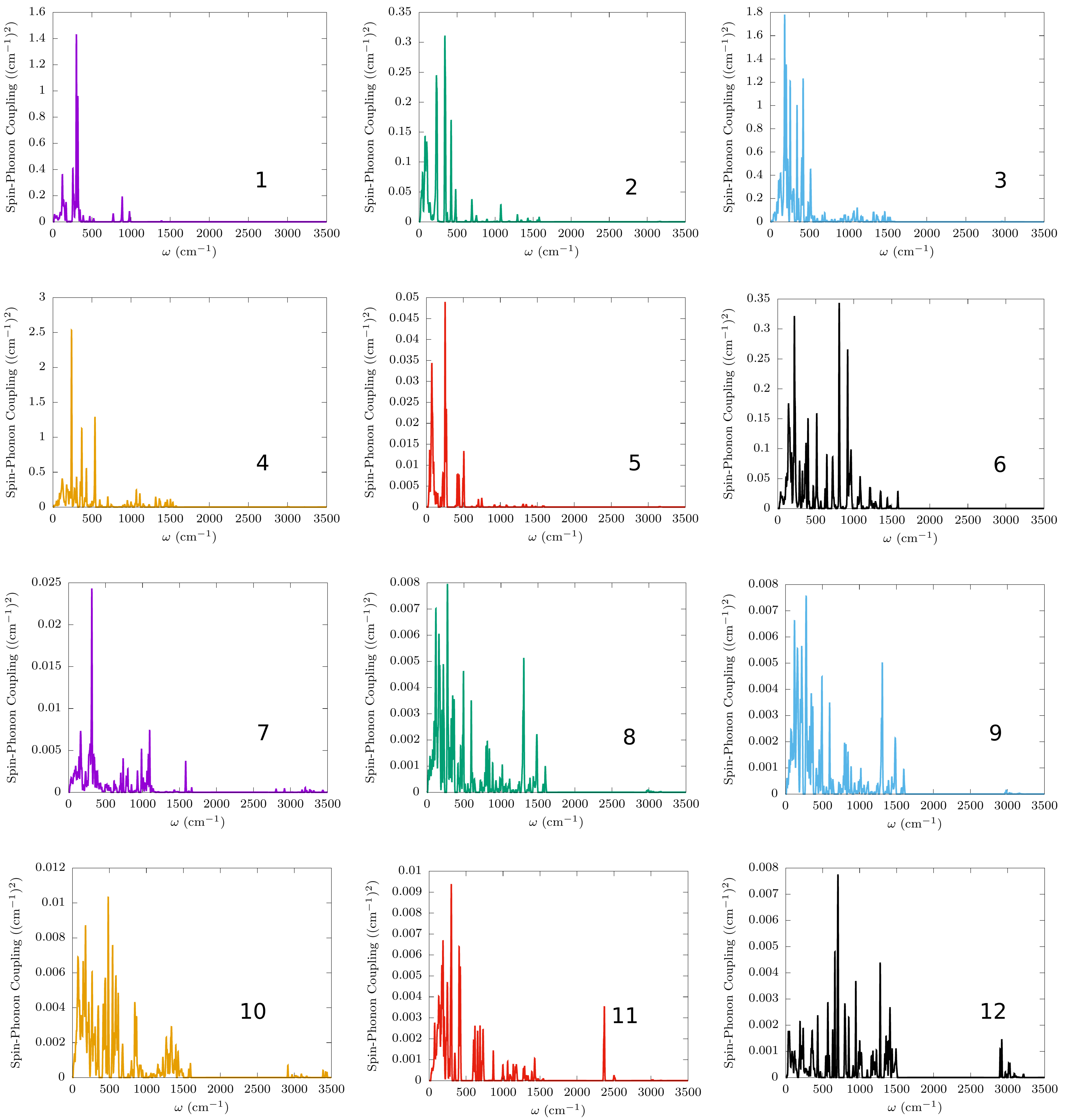}
    \caption*{FIG. S4: \textbf{Spin-Phonon Coupling Density for 1--12.} The spin-phonon coupling density coupling density, $D(\omega)$, has been calculated as $ D(\omega)=\sum_{\alpha}\sum_{ij} \left( \partial \hat{D}_{ij}/\partial Q_{\alpha} \right )^2 \delta(\omega-\omega_{\alpha})$ and $ D(\omega)=\sum_{\alpha}\sum_{lm} \left( \partial \hat{B}^{l}_{m}/\partial Q_{\alpha} \right )^2 \delta(\omega-\omega_{\alpha})$ for Co$^{2+}$ and Dy$^{3+}$, respectively. A Gaussian smearing of 5 cm$^{-1}$ has been applied. Color code for top two panels: violet (\textbf{1}), green (\textbf{2}), blue (\textbf{3}),  yellow (\textbf{4}), red (\textbf{5}), and black (\textbf{6}). Color code for bottom two panels:  violet (\textbf{7}), green (\textbf{8}), blue (\textbf{9}),  yellow (\textbf{10}), red (\textbf{11}), and black (\textbf{12}). }
    \label{S4}
    \end{figure}

\begin{figure}[t]
    \centering
    \includegraphics[scale=1]{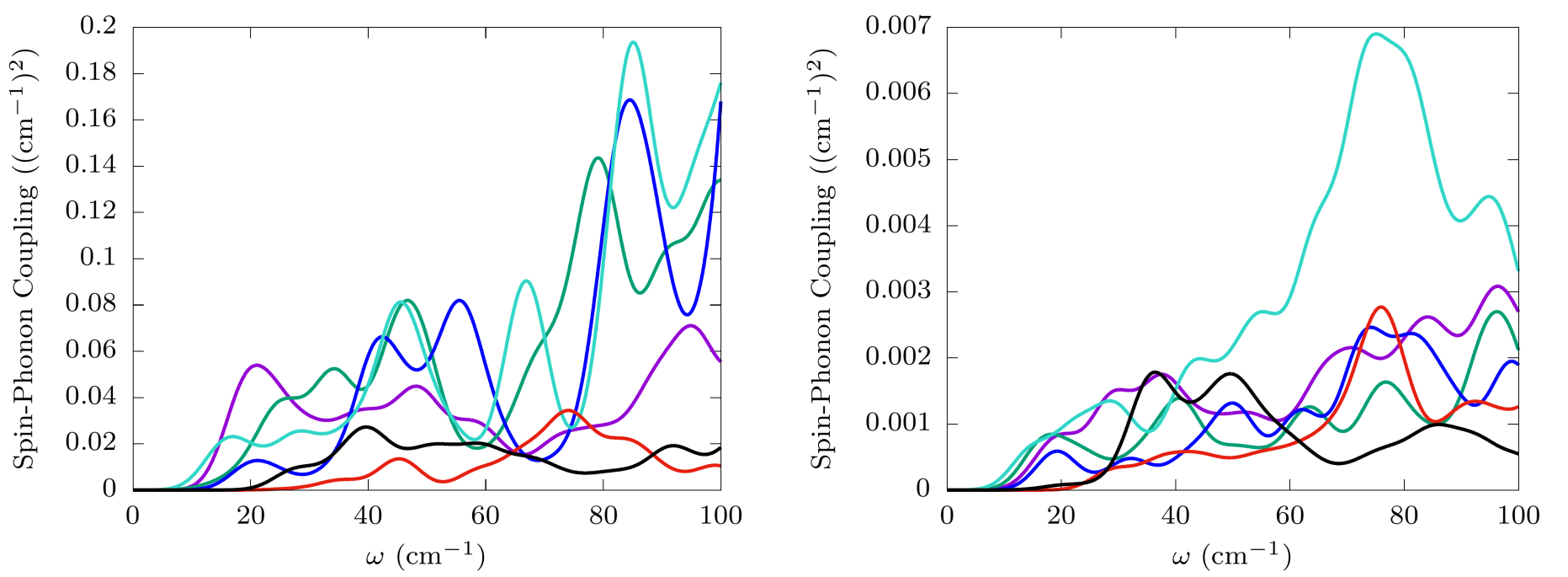}
    \caption*{FIG. S5: \textbf{Comparison of Spin-Phonon Coupling Density for 1--6 (left) and 7-12 (right). } The spin-phonon coupling density coupling density, $D(\omega)$, has been calculated as $ D(\omega)=\sum_{\alpha}\sum_{ij} \left( \partial \hat{D}_{ij}/\partial Q_{\alpha} \right )^2 \delta(\omega-\omega_{\alpha})$ and $ D(\omega)=\sum_{\alpha}\sum_{lm} \left( \partial \hat{B}^{l}_{m}/\partial Q_{\alpha} \right )^2 \delta(\omega-\omega_{\alpha})$ for Co$^{2+}$ and Dy$^{3+}$, respectively. A Gaussian smearing of 5 cm$^{-1}$ has been applied. Color code for left panels: 1 (violet), 2 (green), 3 (blue), 4 (turquoise), 5 (red), and 6 (black). Color code for right panels: 7 (violet), 8 (green), 9 (blue), 10 (turquoise), 11 (red), and 12 (black). }
    \label{S5}
\end{figure}

\begin{figure}[t]
    \centering
    \includegraphics[scale=1]{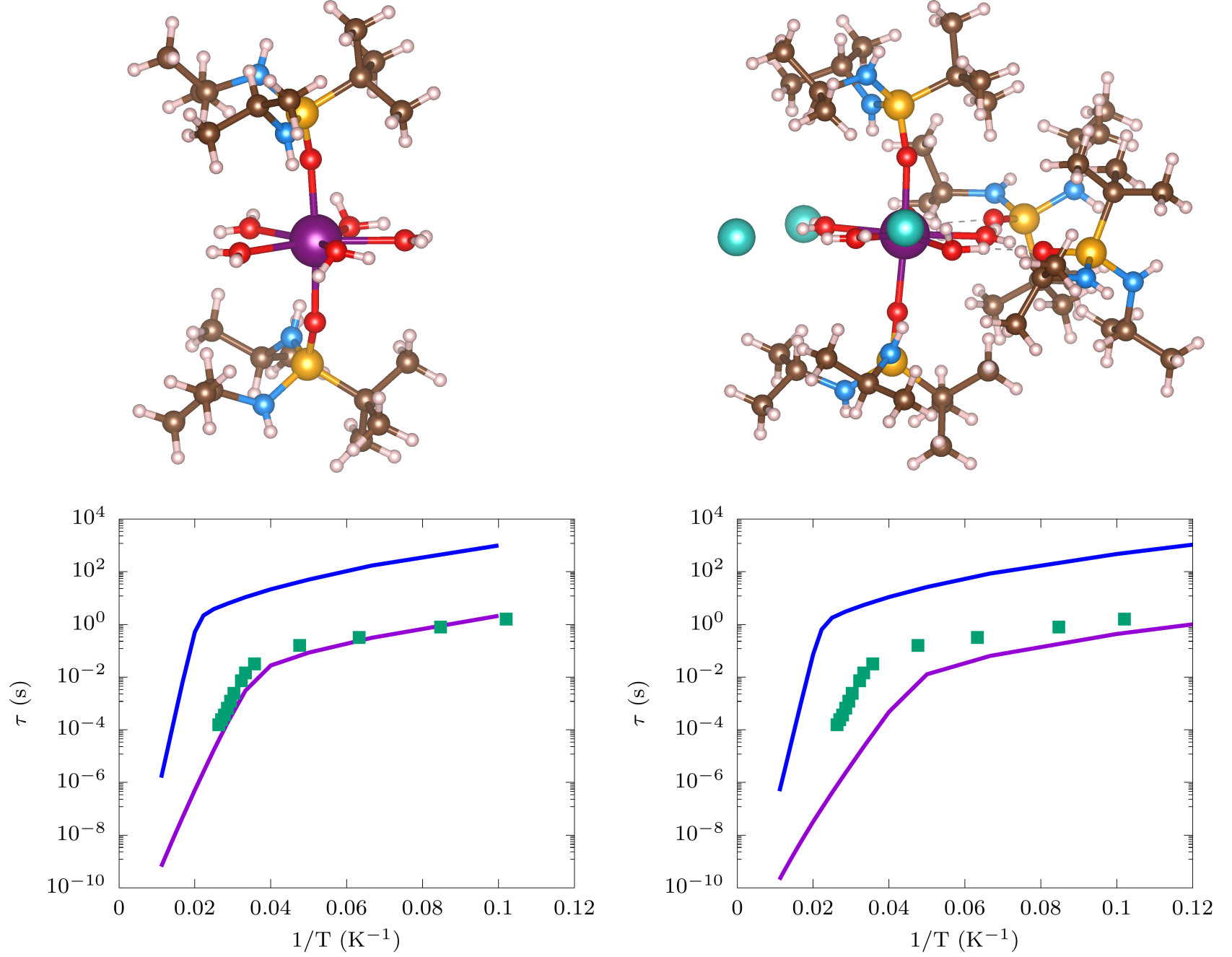}
    \caption*{FIG. S6: \textbf{Simulated values of $\tau$ as a function of $1/T$ for the experimental X-ray coordinates (bottom left) and optimized coordinates (bottom right) of 7.} Blue line represents the simulation result for the model that includes the atoms inside the 1$^{st}$ coordination shell of the \textbf{7} (top left). Similarly, violet line represents simulation results of the model that includes atoms inside 2$^{nd}$ coordination sphere of the \textbf{7} (top right). Symbols corresponds to the experimental values. Color codes for the atoms: Dy in purple, N in light blue, O in red,  I in turquoise, C in dark brown, P in gold, H in pale pink.}
    \label{S6}
\end{figure}

\begin{figure}[t]
    \centering
    \includegraphics[scale=1]{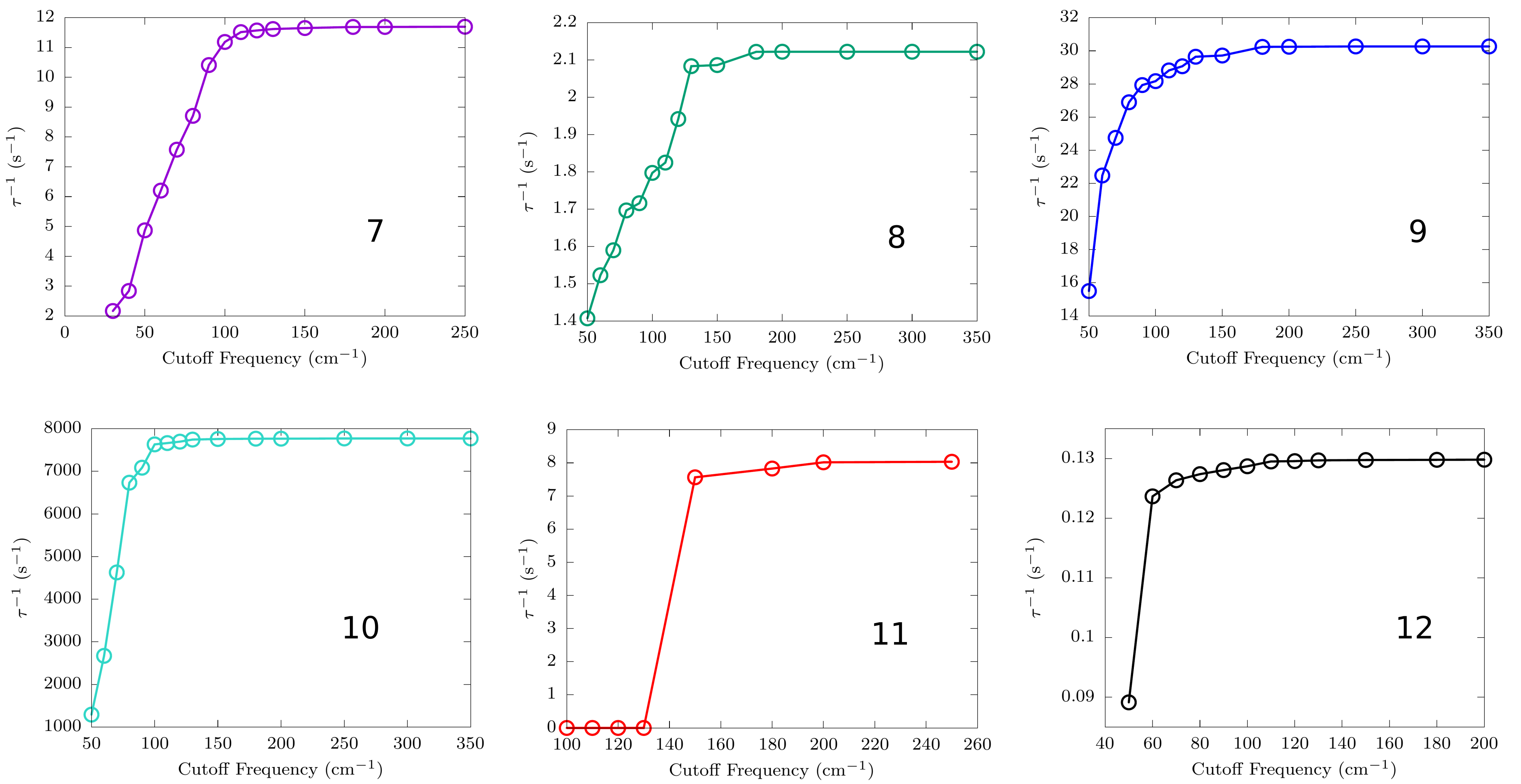}
    \caption*{FIG. S7: \textbf{ Relaxation time as function of phonon energy cutoff.} The Raman relaxation time  $\tau$ for 1 (violet), 2 (green), 3 (blue), 4 (turquoise), 5 (red), and 6 (black) are computed at 20 K including phonons up to a cutoff value of energy}
    \label{S7}
\end{figure}

\clearpage
